\documentclass[twocolumn]{aastex7}
\usepackage[utf8]{inputenc}

\usepackage[multidot]{grffile}

\usepackage[utf8]{inputenc}
\usepackage[T1]{fontenc}
\usepackage{ae,aecompl}
\usepackage{physics}

\DeclareRobustCommand{\VAN}[3]{#2}
\let\VANthebibliography\thebibliography
\def\thebibliography{\DeclareRobustCommand{\VAN}[3]{##3}\VANthebibliography}

\usepackage{graphicx}
\usepackage{amsmath}
\usepackage{amssymb}

\usepackage{comment}
\usepackage{color}
\usepackage{booktabs}
\usepackage{tabularx}
\usepackage{threeparttable}
\usepackage{multirow}
\usepackage{makecell}

\usepackage{academicons}
\usepackage{svg}
\newcommand{\orcid}[1]{\href{https://orcid.org/#1}{\textsuperscript{\includegraphics[width=10pt]{figures/orcid.pdf}}}}

\newcommand{\mmode}[1]{\ifmmode{#1}\else{$#1$}\fi}

\newcommand{\Kepler}[0]{Kepler}

\newcommand{\Teff}[0]{\mmode{T_\text{eff}}}
\newcommand{\Rsun}[0]{\mmode{\text{R}_{\odot}}}
\newcommand{\Msun}[0]{\mmode{\text{M}_{\odot}}}
\newcommand{\Lsun}[0]{\mmode{\text{L}_{\odot}}}
\newcommand{\Rearth}[0]{\mmode{\text{R}_{\earth}}}
\newcommand{\Mearth}[0]{\mmode{\text{M}_{\earth}}}

\newcommand{\logg}[0]{\mmode{\log g}}
\newcommand{\Dnu}[0]{\mmode{\Delta\nu}}
\newcommand{\dnu}[1]{\mmode{\delta\nu_{#1}}}
\newcommand{\numax}[0]{\mmode{\nu_\text{max}}}
\newcommand{\Aosc}[0]{\mmode{A_\text{osc}}}

\newcommand{\muHz}[0]{\mmode{\mu{\rm Hz}}}

\newcommand{\Yinit}[0]{\mmode{Y_{\rm init}}}

\newcommand{\amlt}[0]{\mmode{\alpha_{\rm MLT}}}

\newcommand{\feh}[0]{\mmode{{\rm [Fe/H]}}}
\newcommand{\fbol}[0]{\mmode{{f_{\rm bol}}}}

\newcommand{\logrphk}[0]{\mmode{\log R'_{\rm HK}}}
\newcommand{\Prot}[0]{\mmode{P_{\rm rot}}}
\newcommand{\Pcyc}[0]{\mmode{P_{\rm cyc}}}
\newcommand{\Ro}[0]{\mmode{{\rm Ro}}}

\newcommand{\cyan}[1]{\textcolor{cyan} }

\usepackage{CJKutf8}

\makeatletter
\newcommand\thefontsize[1]{{#1 The current font size is: \f@size pt\par}}
\makeatother

\newcommand{\mass}{0.763}
\newcommand{\emass}{0.020}
\newcommand{\smass}{0.007}

\newcommand{\radius}{0.748}
\newcommand{\eradius}{0.007}
\newcommand{\sradius}{0.002}

\newcommand{\age}{10.151}
\newcommand{\eage}{1.520}
\newcommand{\sage}{0.810}

\begin{document}
\begin{CJK}{UTF8}{gbsn}

\title{K-dwarf Radius Inflation and a 10-Gyr Spin-down Clock \\Unveiled through Asteroseismology of HD~219134 from the Keck Planet Finder}

\newcommand{\PSUAA}{Department of Astronomy \& Astrophysics, 525 Davey Laboratory, The Pennsylvania State University, University Park, PA, 16802, USA}
\newcommand{\PSUCEHW}{Center for Exoplanets \& Habitable Worlds, 525 Davey Laboratory, The Pennsylvania State University, University Park, PA 16802, USA}
\newcommand{\PSETI}{Penn State Extraterrestrial Intelligence Center, 525 Davey Laboratory, The Pennsylvania State University, University Park, PA 16802, USA}
\newcommand{\UA}{Steward Observatory, The University of Arizona, 933 N.\ Cherry Ave, Tucson, AZ 85721, USA}
\newcommand{\Caltech}{Department of Astronomy, California Institute of Technology, Pasadena, CA 91125, USA}
\newcommand{\JHU}{Department of Physics \& Astronomy, Bloomberg Center, Johns Hopkins University, Baltimore, MD 21218, USA}
\newcommand{\Macquarie}{School of Mathematical and Physical Sciences, Macquarie University, Balaclava Road, North Ryde, NSW 2109, Australia}
\newcommand{\CUBoulder}{Department of Physics, 390 UCB, University of Colorado, Boulder, CO 80309, USA}
\newcommand{\JPL}{Jet Propulsion Laboratory, California Institute of Technology, 4800 Oak Grove Drive, Pasadena, CA 91109, USA}
\newcommand{\MITEAPS}{Department of Earth, Atmospheric, and Planetary Sciences, Massachusetts Institute of Technology, Cambridge, MA 02139, USA}
\newcommand{\MITKavli}{Kavli Institute for Astrophysics and Space Research, Massachusetts Institute of Technology, Cambridge, MA 02139, USA}
\newcommand{\UCI}{Department of Physics \& Astronomy, The University of California, Irvine, Irvine, CA 92697, USA}
\newcommand{\Carnegie}{Earth and Planets Laboratory, Carnegie Institution for Science, 5241 Broad Branch Road, NW, Washington, DC 20015, USA}
\newcommand{\PSUICS}{Institute for Computational and Data Sciences, The Pennsylvania State University, University Park, PA 16802, USA}
\newcommand{\PSUCASt}{Center for Astrostatistics, 525 Davey Laboratory, The Pennsylvania State University, University Park, PA 16802, USA}
\newcommand{\Princeton}{Department of Astrophysical Sciences, Princeton University, 4 Ivy Lane, Princeton, NJ 08540, USA}
\newcommand{\IAS}{Institute for Advance Study, 1 Einstein Drive, Princeton, NJ 08540, USA}
\newcommand{\Tsinghua}{Department of Astronomy, Tsinghua University, Beijing 100084, China}
\newcommand{\FlatironCCA}{Center for Computational Astrophysics, Flatiron Institute, 162 Fifth Avenue, New York, NY 10010, USA}
\newcommand{\ETH}{ETH Zurich, Institute for Particle Physics \& Astrophysics, Zurich, Switzerland}
\newcommand{\UCO}{UC Observatories, University of California, Santa Cruz, CA 95064, USA}
\newcommand{\SantaCruz}{University of California, Santa Cruz}
\newcommand{\WMKO}{W.\ M.\ Keck Observatory, 65-1120 Mamalahoa Hwy, Kamuela, HI 96743, USA}
\newcommand{\SSL}{Space Sciences Laboratory, University of California, Berkeley, CA 94720, USA}
\newcommand{\UH}{Institute for Astronomy, University of Hawai‘i, 2680 Woodlawn Drive, Honolulu, HI 96822, USA}
\newcommand{\UCB}{Department of Astronomy, 501 Campbell Hall, University of California, Berkeley, CA 94720, USA}
\newcommand{\UCLA}{Department of Physics \& Astronomy, University of California Los Angeles, Los Angeles, CA 90095, USA}
\newcommand{\nexsci}{NASA Exoplanet Science Institute/Caltech-IPAC, California Institute of Technology, Pasadena, CA
91125, USA}
\newcommand{\COO}{Caltech Optical Observatories, California Institute of Technology, Pasadena, CA 91125, USA}
\newcommand{\Sydney}{Sydney Institute for Astronomy (SIfA), School of Physics, University of Sydney, NSW 2006, Australia}
\newcommand{\Kansas}{Department of Physics and Astronomy, University of Kansas, Lawrence, KS, USA}
\newcommand{\Warwick}{Physics Department, University of Warwick, Coventry CV4 7AL, United Kingdom}
\newcommand{\Yale}{Department of Astronomy, Yale University, New Haven, CT 06511, USA}
\newcommand{\ICL}{Astrophysics Group, Department of Physics, Imperial College London, Prince Consort Rd, London SW7 2AZ, UK}
\newcommand{\Schmidt}{Astrophysics \& Space Institute, Schmidt Sciences, New York, NY 10011, USA}
\newcommand{\Amsterdam}{University of Amsterdam}
\newcommand{\ND}{Department of Physics and Astronomy, University of Notre Dame, Notre Dame, IN 46556, USA}
\newcommand{\Geneva}{Observatoire Astronomique de l'Université de Genève, Chemin Pegasi 51, 1290 Versoix, Switzerland}
\newcommand{\PortoIAC}{Instituto de Astrof\'{i}sica e Ci\^{e}ncias do Espaço, Universidade do Porto, Rua das Estrelas, 4150-762 Porto, Portugal}
\newcommand{\Porto}{Departamento de F\'{i}sica e Astronomia, Faculdade de Ci\^{e}ncias da Universidade do Porto, Rua do Campo Alegre, s/n, 4169-007 Porto, Portugal}
\newcommand{\Aarhus}{Stellar Astrophysics Centre (SAC), Department of Physics and Astronomy, Aarhus University, Ny Munkegade 120, 8000 Aarhus C, Denmark}


\author[0000-0003-3020-4437]{Yaguang~Li (李亚光)}
\affiliation{\UH}
\email[show]{yaguangl@hawaii.edu}

\author[orcid=0000-0001-8832-4488]{Daniel Huber}
\email{huberd@hawaii.edu}
\affiliation{\UH}

\author[orcid=0000-0001-7664-648X]{J. M. Joel Ong (王加冕)}
\email{joelong@hawaii.edu}
\altaffiliation{NASA Hubble Fellow}
\affiliation{\UH}

\author[orcid=0000-0002-4284-8638]{Jennifer van Saders}
\email{jlvs@hawaii.edu}
\affiliation{\UH}



\author[orcid=0009-0008-6039-6381]{R.R. Costa}
\email{up201909641@edu.fc.up.pt}
\affiliation{\PortoIAC}
\affiliation{\Porto}

\author[orcid=0009-0006-0423-2353]{Jens Reersted Larsen}
\email{jensrl@phys.au.dk}
\affiliation{\Aarhus}

\author[orcid=0000-0002-6163-3472]{Sarbani Basu}
\email{sarbani.basu@yale.edu}
\affiliation{\Yale}

\author[orcid=0000-0001-5222-4661]{Timothy R. Bedding}
\email{tim.bedding@sydney.edu.au}
\affiliation{\Sydney}


\author[orcid=0000-0002-8958-0683]{Fei Dai (戴飞)}
\email{fdai@hawaii.edu}
\affiliation{\UH}


\author[orcid=0000-0003-1125-2564]{Ashley Chontos}
\email{ashleychontos@astro.princeton.edu}
\affiliation{\Princeton}

\author[orcid=0000-0001-6416-1274]{Theron W. Carmichael}
\email{tcarmich@hawaii.edu}
\affiliation{\UH}

\author[orcid=0000-0003-3244-5357]{Daniel Hey}
\email{dhey@hawaii.edu}
\affiliation{\UH}




\author[orcid=0000-0002-9037-0018]{Hans Kjeldsen}
\email{hans@phys.au.dk}
\affiliation{\Aarhus}

\author[orcid=0000-0003-2400-6960]{Marc Hon}
\email{mtyhon@mit.edu}
\affiliation{\MITKavli}


\author[orcid=0000-0002-4588-5389]{Tiago L. Campante}
\email{Tiago.Campante@astro.up.pt}
\affiliation{\PortoIAC}
\affiliation{\Porto}

\author[orcid=0000-0003-0513-8116]{Mário J. P. F. G. Monteiro}
\email{mario.monteiro@astro.up.pt}
\affiliation{\PortoIAC}
\affiliation{\Porto}

\author[orcid=0000-0002-8661-2571]{Mia Sloth Lundkvist}
\email{lundkvist@phys.au.dk}
\affiliation{\Aarhus}

\author[orcid=0000-0003-2657-3889]{Nicholas Saunders}
\email{saunders.nk@gmail.com}
\altaffiliation{NSF Graduate Research Fellow}
\affiliation{\UH}







\author[orcid=0000-0002-0531-1073]{Howard Isaacson}
\email{hisaacson@berkeley.edu}
\affiliation{\UCB}

\author[orcid=0000-0001-8638-0320]{Andrew W. Howard}
\email{ahoward@caltech.edu}
\affiliation{\Caltech}

\author[orcid=0009-0004-4454-6053]{Steven R.\ Gibson}
\email{sgibson@caltech.edu}
\affiliation{\COO}

\author[orcid=0000-0003-1312-9391]{Samuel Halverson}
\email{samuel.halverson@jpl.nasa.gov}
\affiliation{\JPL}

\author{Kodi Rider}
\email{kodi.rider@ssl.berkeley.edu}
\affiliation{\SSL}

\author[orcid=0000-0001-8127-5775]{Arpita Roy}
\email{arpita308@gmail.com}
\affiliation{\Schmidt}

\author[orcid=0000-0002-6525-7013]{Ashley D.\ Baker}
\email{abaker@caltech.edu}
\affiliation{\COO}

\author[orcid=0009-0002-2419-8819]{Jerry Edelstein}
\email{jerrye@ssl.berkeley.edu}
\affiliation{\SSL}

\author{Chris Smith}
\email{christopher.smith@berkeley.edu}
\affiliation{\SSL}

\author[orcid=0000-0003-3504-5316]{Benjamin J.\ Fulton}
\email{bjfulton@ipac.caltech.edu}
\affiliation{\nexsci}

\author[orcid=0000-0002-6092-8295]{Josh Walawender}
\email{jwalawender@keck.hawaii.edu}
\affiliation{\WMKO}

\begin{abstract}
We present the first asteroseismic analysis of the K3\,V planet host HD~219134, based on four consecutive nights of radial velocities collected with the Keck Planet Finder. We apply Gold deconvolution to the power spectrum to disentangle modes from sidelobes in the spectral window, and extracted 25 mode frequencies with spherical degrees $0\leq\ell\leq3$. We derive the fundamental properties using five different evolutionary-modeling pipelines and report
a mass of \mass{} $\pm$ \emass{} (stat) $\pm$ \smass{} (sys) \Msun{}, a radius of \radius{} $\pm$ \eradius{} (stat) $\pm$ \sradius{} (sys) \Rsun{}, and an age of \age{} $\pm$ \eage{} (stat) $\pm$ \sage{} (sys) Gyr. Compared to the interferometric radius 0.783 $\pm$ 0.005~\Rsun{}, the asteroseismic radius is 4\% smaller at the 4-$\sigma$ level --- a discrepancy not easily explained by known interferometric systematics, modeling assumptions on atmospheric boundary conditions and mixing lengths, magnetic fields, or tidal heating. HD~219134 is the first main-sequence star cooler than 5000~K with an asteroseismic age estimate and will serve as a critical calibration point for stellar spin-down relations. We show that existing calibrated prescriptions for angular momentum loss, incorporating weakened magnetic braking with asteroseismically constrained stellar parameters, accurately reproduce the observed rotation period. 
Additionally, we revised the masses and radii of the super-Earths in the system, which support their having Earth-like compositions. Finally, we confirm that the oscillation amplitude in radial velocity scales as $(L/M)^{1.5}$ in K dwarfs, in contrast to the $(L/M)^{0.7}$ relation observed in G dwarfs. These findings provide significant insights into the structure and angular momentum loss of K-type stars.
\end{abstract}


\section{Introduction} \label{sec:intro}

Asteroseismology, the study of stellar oscillations, is a powerful method for measuring fundamental stellar properties and studying interior dynamics. Early detections of oscillations in solar-type stars---see \citet{Bedding+Kjeldsen2003} for a review---were made possible through ground-based radial velocity (RV) observations, such as those observed in Procyon \citep{brown1991-procyon, mosser1998-procyon, Martic1999}, $\beta$~Hyi \citep{Bedding2001} and $\alpha$~Cen~A \citep{Bouchy+Carrier2001, Bouchy+Carrier2002}. 
These observations confirmed the presence of solar-like oscillations in cool stars beyond the Sun, which are stochastically excited and damped by surface turbulence \citep{goldreich1977-excitation}.

In stars with solar-like oscillations, acoustic modes are observed around the frequency of maximum power, \numax{}. Their non-rotating frequencies are well approximated by the asymptotic relation involving two quantum numbers, the radial order, $n$, and angular degree, $\ell$ \citep{tassoul1980-asymp,scherrer_detection_1983}:
\begin{equation}\label{eq:asymp}
    \nu_{n,\ell} \simeq \Delta\nu \left( n + \frac{\ell}{2} + \epsilon \right) - \delta\nu_{0,\ell}.
\end{equation}
The large frequency separation, \Dnu{}, depends primarily on the sound travel time across the star and, to first order, scales with the square root of the mean stellar density \citep{Ulrich1986}. 
The phase offset, $\epsilon$, generally falls between 0.8 and 1.6 in main-sequence stars \citep{white2012}, and depends on the physical structure of the star, primarily as concentrated near the center and surface \citep{roxburgh2003-ratio}. 
The small separations, \dnu{0,\ell}, quantify the frequency offsets between modes of different $\ell$ but with same $n$. In main-sequence stars, \dnu{0,2} is sensitive to the chemical gradient near the core \citep{tassoul_asymptotic_1994,roxburgh_asymptotic_1994}, making it a useful diagnostic for tracing hydrogen-burning processes and, therefore, for determining stellar ages on the main sequence \citep{jcd1984-02,white2011-cd,hon2024-flow}.

Since the advent of space missions in the 2000s, photometric measurements have become a primary method for detecting stellar oscillations, with the significant advantage of allowing simultaneous observations of many stars.
Missions like WIRE \citep{Buzasi2002}, MOST \citep{Matthews2000}, CoRoT \citep{Auvergne2009}, Kepler \citep{Borucki2010} and TESS \citep{Ricker2015} have substantially increased the number of detections of solar-like oscillations in G- and F-type main-sequence stars, as well as in sub- and red giants spanning a wide range of evolutionary stages \citep[see reviews by][]{Chaplin+Miglio2013, Jackiewicz2021}.  
The high precision of these photometric observations allows for detailed analysis of oscillation modes, probing the structure and dynamics of stellar interiors, including rotation, magnetism, and binary merging \citep[see][for a review]{aerts2021-review}. 

Recent significant advances in Doppler precision, driven by the search for Earth-like planets, have opened new opportunities for detecting stellar oscillations with much lower amplitudes. 
Extreme Precision Radial Velocity (EPRV) techniques now achieve radial velocity measurements with precisions in the range of 10--30 cm/s \citep{Fischer++2016, Wright2018}.
This level of precision makes it possible to detect the intrinsic oscillations of K dwarfs, which have amplitudes ranging from 1 to 6 cm/s. 
These oscillations are very challenging to detect in photometry due to the presence of granulation noise, which is significantly lower (relative to the oscillations) in RV measurements from spatially-averaged line profiles \citep{Harvey1988,Grundahl++2007,Asplund2000,Kjeldsen+Bedding2011}. 

\begin{figure}
    \centering
    \includegraphics{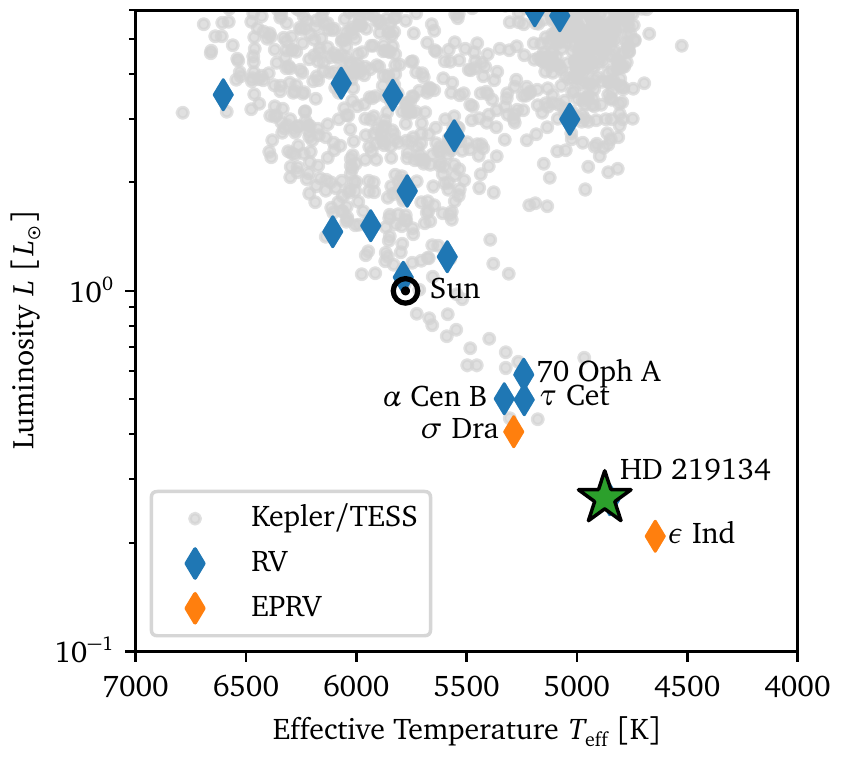}
    \caption{H--R diagram showing the footprint of asteroseismology conducted using space-based photometry \citep[e.g.][]{Chaplin2015, Campante++2015,Lund2016,lund2017-legacy,serenelli2017-apokasc,yu2018-rg,liyg2020,hatt2023-tess,Gonzalez-Cuesta2023,zhou2024-tess,Lund2024} and ground-based radial velocity (RV) measurements \citep[e.g.][]{kjeldsen2005-acenb, Carrier+Eggenberger2006,Teixeira++2009}. The diagram highlights recent contributions made possible by Extreme Precision Radial Velocity (EPRV) instruments \citep{campante2024-eps-indi,hon2024-sig-dra}, which have significantly expanded the parameter space for K dwarf studies.}
    \label{fig:hrd}
\end{figure}

Until recently, the only K dwarfs with oscillations detected in RV were the bright K1\,V star $\alpha$~Cen~B \citep{Carrier+Bouchy2003, kjeldsen2005-acenb} and the slightly hotter K0\,V star 70~Oph~A \citep{Carrier+Eggenberger2006}. 
The arrival of EPRV instruments has ushered in a new era of K-dwarf asteroseismology
Oscillations in the K5\,V star $\epsilon$ Indi were reported by \citet{campante2024-eps-indi} with six half-nights of observations with ESPRESSO on the VLT \citep{pepe2021-espresso}, and also at lower signal-to-noise with HARPS and UVES by \citet{lundkvist2024-eps-ind}.  
\citet{hon2024-sig-dra} detected oscillations in the K0\,V star $\sigma$ Draconis using one night of observations with the Keck Planet Finder (KPF) on the Keck~I telescope \citep{gibson2016-kpf,gibson2018-kpf,gibson2020-kpf,gibson2024-kpf}. 
Although TESS photometryalso shows oscillations in $\sigma$ Draconis from 14 sectors of data, the KPF RV power spectrum revealed oscillation modes over a broader frequency range due to the higher signal-to-noise ratio achieved in the RV data.
Figure~\ref{fig:hrd} illustrates the expanding asteroseismology footprint, with EPRV techniques populating the K dwarf parameter space.

While a single night of data is sufficient to detect both \numax{} and \Dnu{} (provided sufficient SNR is reached), achieving the resolution required to measure small frequency spacings like \dnu{0,2} remains challenging. 
For example, an 8-hour time series only provides a frequency resolution of $\approx$34~\muHz{}, which exceeds the typical \dnu{0,2} values of about 10~\muHz{} for main-sequence stars \citep[e.g.][]{white2011-cd}.
We can address this limitation using multiple nights of data.
For this paper, we have obtained four nights of data collected with the KPF, corresponding to a frequency resolution of $\approx$3~\muHz{}. We report the detection of oscillations in HD~219134 and the determination of its benchmark mass, radius, and age via asteroseismology.

\subsection{The K3\,V exoplanet host HD~219134}

HD~219134 (HR~8832) is a nearby K3\,V dwarf ($d=6.5$~pc; $V=5.57$). It has an angular diameter precisely measured through interferometry \citep{ligi2019-219134,huber2016-itf,elliott2025}, enabling an accurate determination of its radius (see \S\ref{subsec:obs}). Asteroseismology provides an independent method to estimate the stellar radius by probing the star’s interior through oscillation frequencies. This star presents an opportunity to directly compare these two techniques.

HD~219134 has a rotation period of $\approx$42 days \citep{motalebi2015-219134,folsom2018-zdi}. 
Cool stars experience gradual spin-down over time, making rotation period a valuable proxy for stellar age --- the principle underpinning gyrochronology.
However, the applicability of the rotation-age relation relies on calibrations using stars with well-known ages, primarily those in open clusters \citep{curtis2020-147}. 
Importantly, HD~219134 rotates more slowly than all stars of similar spectral type in the current open cluster calibrators.
Furthermore, recent studies of K-dwarf stars have revealed a phase of apparently stalled spin-down \citep{curtis2019-6811,curtis2020-147,agueros2018-752}. It remains unclear for how long the epoch of stalling persists or the degree to which it affects the rotation rates of old stars.
Measuring the age of this star using asteroseismology would provide a valuable test of the rotation-age spin-down relationship for stars at advanced ages.

HD~219134 hosts five known RV planets, including the transiting super-Earths HD~219134~b and~c, whose orbital periods are 3.09 and 6.87 days \citep{vogt2015-219134,motalebi2015-219134,seager2021-219134,rosenthal2021-cls,gillon2017-219134,kokori2023-exoclock}. A detailed seismic characterization of the host star's mass and radius would refine the properties of these planets.

The paper is organized as follows. We report our detection of oscillations and extract oscillation frequencies (\S\ref{sec:obs}). 
We derive stellar properties using asteroseismic modeling (\S\ref{sec:modeling}). 
We highlight the discrepancy between our asteroseismic radius and an interferometric radius and explore potential explanations (\S\ref{sec:radius}).
We discuss the implication of stellar age for gyrochronology by testing models with angular momentum loss (\S\ref{sec:rotation}). 
We revise planet properties using the new stellar mass and radius (\S\ref{sec:planet}). 
Finally, we analyze mode amplitudes (\S\ref{sec:modes}), and present our conclusions (\S\ref{sec:conc}).

\begin{figure*}
    \centering
    \includegraphics{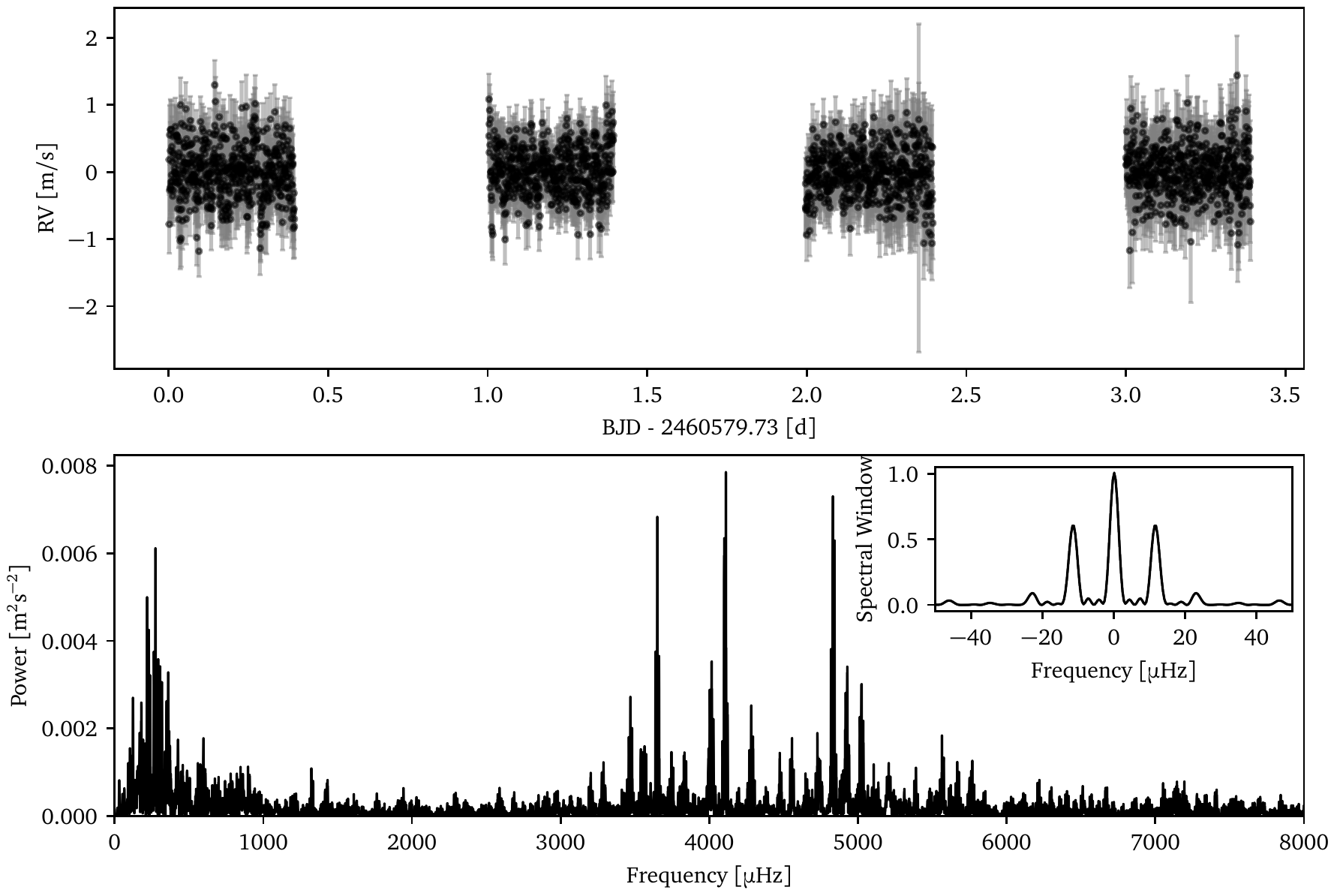}
    \caption{Asteroseismic observations of HD~219134 over four consecutive nights using the Keck Planet Finder. Top: Radial-velocity time series after filtering out signals with periods longer than 1.2 hours. Bottom: Power spectrum of the RV time series, weighted by the reported RV uncertainties, displaying a clear power excess around 4500~\muHz{}. The inset shows the spectral window.}
    \label{fig:data}
\end{figure*}

\begin{figure}
    \centering
    \includegraphics{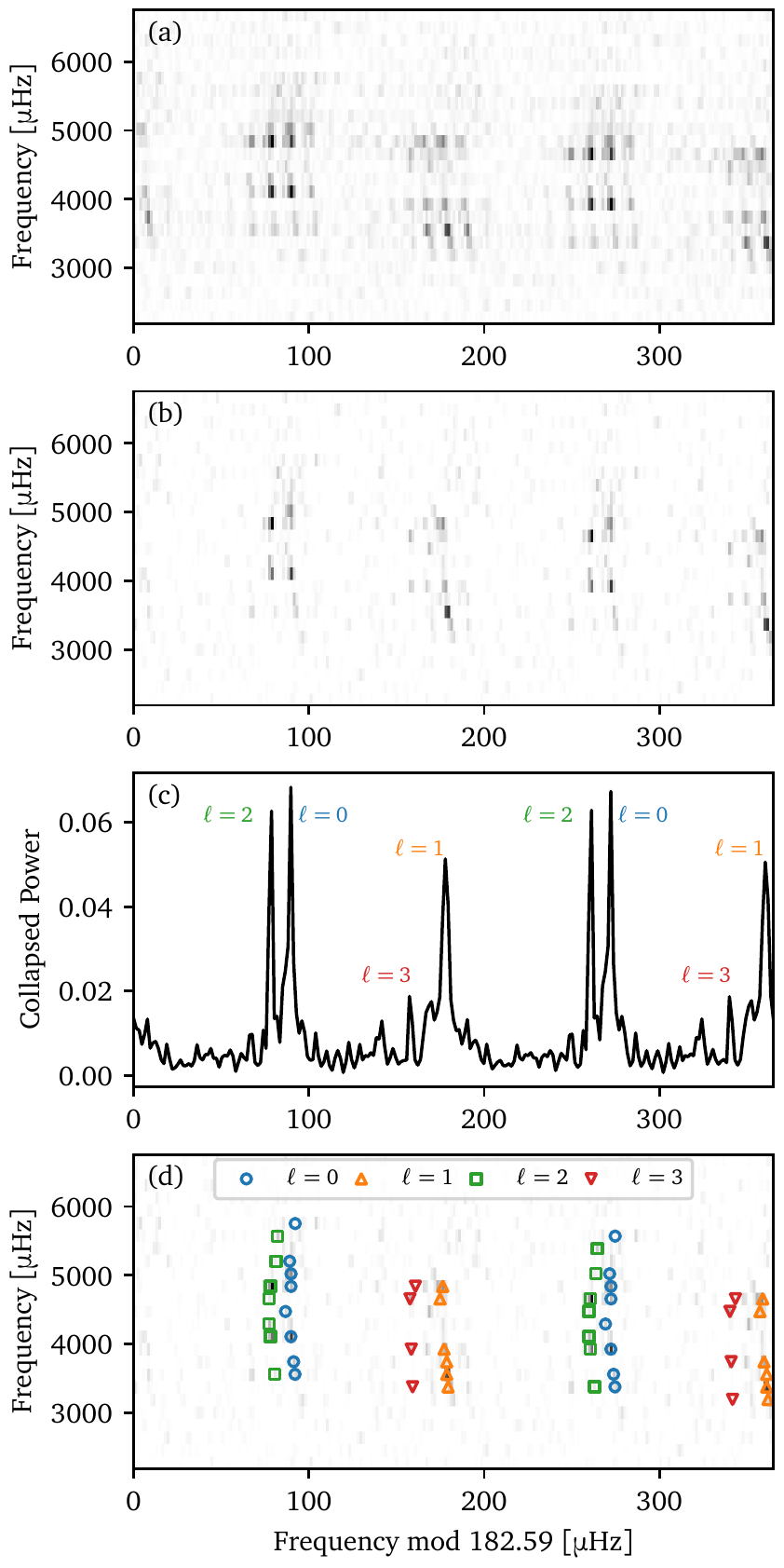}
    \caption{\'Echelle diagrams showing structures of regular frequency spacings. Panel a: replicated \'echelle diagram of the original power spectrum. Panel b: \'echelle diagram of the power spectrum, deconvolved from the spectral window and then convolved with a 1~$\mu$Hz width Gaussian filter for clarity. Panel c: collapsed \'echelle diagram, summing the power along the y-axis of panel b. Panel d: same as panel b, but highlighting the extracted oscillation modes. }
    \label{fig:echelle}
\end{figure}

\section{Oscillation Frequencies}\label{sec:obs}

\begin{deluxetable}{cccccccc}
\tabletypesize{\footnotesize}
\tablecolumns{4}
\tablewidth{\textwidth} 
\tablecaption{ Oscillation frequencies of HD~219134, including corrections for line-of-sights Doppler shifts. \label{table:freqs}}
\tablehead{
\colhead{\hspace{0.4cm}$n$}\hspace{0.4cm} & \colhead{\hspace{0.4cm}$\ell$}\hspace{0.4cm} & \colhead{\hspace{0.4cm}$\nu_{n,\ell}$ (\muHz{})}\hspace{0.4cm} &  \colhead{\hspace{0.4cm}$\sigma(\nu_{n,\ell})$ (\muHz{})}\hspace{0.4cm} \\
}
\startdata
18 & 0 & 3561.03 & 0.58 \\
19 & 0 & 3742.84 & 1.07 \\
21 & 0 & 4106.45 & 0.64 \\
23 & 0 & 4468.50 & 0.91 \\
25 & 0 & 4836.80 & 0.95 \\
26 & 0 & 5019.39 & 0.85 \\
27 & 0 & 5201.19 & 0.95 \\
30 & 0 & 5752.07 & 1.03 \\
17 & 1 & 3465.84 & 0.84 \\
18 & 1 & 3647.65 & 0.43 \\
19 & 1 & 3830.23 & 0.90 \\
20 & 1 & 4011.26 & 0.65 \\
24 & 1 & 4739.26 & 0.95 \\
25 & 1 & 4923.41 & 0.64 \\
17 & 2 & 3549.33 & 0.62 \\
20 & 2 & 4094.75 & 0.85 \\
21 & 2 & 4276.55 & 1.05 \\
23 & 2 & 4641.73 & 1.00 \\
24 & 2 & 4825.09 & 0.52 \\
26 & 2 & 5193.39 & 1.12 \\
28 & 2 & 5559.34 & 0.98 \\
16 & 3 & 3445.55 & 1.05 \\
19 & 3 & 3992.53 & 1.07 \\
23 & 3 & 4722.10 & 0.93 \\
24 & 3 & 4907.80 & 1.02 \\
\enddata
\end{deluxetable}


We obtained approximately 9 hours of observations per night over 4 consecutive nights, from September 25 to 28, 2024 (HST), using the Keck Planet Finder (KPF) on the Keck~I telescope \citep{gibson2016-kpf,gibson2018-kpf,gibson2020-kpf,gibson2024-kpf}. A total of 2,155 radial velocity measurements were collected. 
We used 45-second exposures, with a readout time of 15 seconds, resulting in a cadence of 1 minute per observation.
Observing conditions were excellent and stable, with seeing ranging from 0.4’’ to 0.8’’. The signal-to-noise ratio (SNR) of the spectra at 747~nm remained consistently around 500. 

The spectra were processed with KPF Data Reduction Pipeline \citep[DRP;][]{gibson2020-kpf}\footnote{\url{https://github.com/Keck-DataReductionPipelines/KPF-Pipeline?tab=readme-ov-file}}. 
KPF DRP executes several processing steps, including quadrant stitching, flat-field correction, order tracing, optimal spectral extraction, and wavelength calibration.
RVs are determined from a cross-correlation function (CCF) mask method, using the same K2 binary CCF mask as in the ESPRESSO pipeline \citep{pepe2021-espresso}. A mean position of the CCF is fitted using a Gaussian plus top-hat function. RVs are computed separately for each of the three KPF slices and for each camera (green and red). The RVs from individual slices are combined through a weighted average, where the weights are inversely proportional to the photon-limited RV uncertainties for each slice. The RVs from the cameras are adjusted by median subtraction and corrected for instrumental drift, based on laser frequency comb spectra obtained before and after the stellar RV observations. The drift-corrected RV time series for the two cameras are then combined into a single dataset using a flux-weighted mean.

To calculate the power spectrum of the RV time series, we applied the Lomb-Scargle method \citep{lomb1976,scargle1982}, using the reported RV uncertainties as weights \citep{Frandsen++1995}.
Normalization of the power spectrum followed the convention of \citet{kb95}: a sine wave with amplitude $A$ produces a peak with a height of $A^2$ in the power spectrum, and $A^2T_{\rm obs}$ in the power density spectrum, with $T_{\rm obs}$ being the effective observing duration.
In Figure~\ref{fig:data}, we show the RV time series, along with the power spectrum.

The power spectrum in the high-frequency regime contains only white noise, from which we estimated the time-domain RV scatter, $\sigma_{\rm noise}$, due to photon and instrumental noise. The scatter was calculated as $\sigma_{\rm noise} = \sqrt{P_{\rm mean}/(2\Delta t)}$, where $P_{\rm mean}$ is the mean power density between 6500~\muHz{} and 8000~\muHz{}, and $\Delta t$ is the cadence. We obtained a value of time-domain scatter at 0.37 m/s, which is in agreement with the average RV uncertainties reported from the reduction pipeline at 0.41 m/s.

Regular data gaps in time domain inevitably cause multiple peaks near the true frequency in the Fourier domain, a phenomenon known as spectral leakage. The observed power spectrum is the convolution of the true spectrum (as from a continuous and infinite time series) with the spectral window function. To examine the shape of the spectral window for our dataset, we generate sine and cosine waves at an arbitrary test frequency, sampled at the timestamps of our RV data. We then calculate and average the power spectra of both waves, which helps reduce edge effects from signal apodization. The inset of Figure~\ref{fig:data} shows the spectral window centred around the test frequency. The strongest sidelobes appear at multiples of 1~c/d ($\approx$11.57~\muHz{}), due to the existence of daily gaps.

The daily sidelobes present a significant challenge in identifying the \dnu{0,2} spacing, as it is very close to 1~c/d in K-type main-sequence stars \citep{white2011-cd}. 
This can be seen from Figure~\ref{fig:echelle}(a), which shows the power spectrum folded into segments of \Dnu{} and displayed in an \'echelle format. In such diagrams, modes of the same spherical degree $\ell$ form approximately vertical ridges from consecutive radial orders $n$, as predicted by the asymptotic relation (Equation~\ref{eq:asymp}). 
Based on observations and theoretical predictions of the phase offset $\epsilon$ \citep{white2011-cd, Ong2019-eps}, we expect $\epsilon$ to be approximately 1.4 given the \Dnu{} value of HD~219134. This implies that the ridges in Figure~\ref{fig:echelle}(a) at an abscissa of $\approx$90~\muHz{} corresponds to the $\ell=0$ and $\ell=2$ modes, while the ridge at 180~\muHz{} corresponds to the $\ell=1$ (and $\ell=3$, if present) modes. 
Due to geometric cancellation, only oscillation modes with low spherical degrees ($0\leq\ell\leq3$) are expected to be visible. 
However, Figure~\ref{fig:echelle}(a) shows more than four vertical ridges, many of which are sidelobes --- replicas of true ridges with offsets at multiples of 1~c/d. In particular, the true $\ell=2$ ridge could overlap with the left sidelobe of the $\ell=0$ ridge, due to the closeness of $\dnu{0,2}$ to 1~c/d.

To better reveal the frequency structure, we deconvolved the power spectrum against the spectral window using the Gold algorithm \citep{gold1964}, by applying the numerical implementation described by \citet{morhac2003-gold}. The algorithm takes the power spectrum and the spectral window as inputs, and iteratively solves for the underlying true spectrum with some regularization\footnote{\url{https://github.com/parallelpro/maemae}}. 
We selected this algorithm over other deconvolution schemes for its desirable properties: (1) it ensures the solution remains positive when both the input data and spectral window are positive; and (2) it provides non-oscillating solutions for isolated peaks, effectively suppressing ringing artifacts \citep{morhac2009-deconv}. 
A quantitative evaluation of its performance in asteroseismic applications will be presented in a forthcoming paper; in this work, we applied it solely as a visual aid to mode identification, but otherwise applied traditional fitting techniques to derive estimates and uncertainties of mode frequencies.
Figure~\ref{fig:echelle}(b) shows the spectrum after deconvolution, revealing clear $\ell=0$--$3$ mode ridges. These are more apparent in Figure~\ref{fig:echelle}(c), which shows the collapsed power spectrum from summing power along the vertical direction in Figure~\ref{fig:echelle}(b).

From Figure~\ref{fig:echelle}(b) and (c), we can identify the peaks along each ridge, which we preliminarily considered as candidate oscillation modes. 
We extracted their frequencies by simultaneously fitting sine waves at these frequencies to the RV time series, obtaining the corresponding amplitudes and phases. Signals with amplitudes exceeding 3.5 times the noise level (0.37 m/s) were retained as the final list of modes.
These are highlighted in Figure~\ref{fig:echelle}(d).

Furthermore, we corrected the line-of-sight Doppler shifts for these frequencies following \citet{davies2014-los}, using a radial velocity of $-18368$~m/s based on an average value over four nights. This adjustment reduced the frequencies by an average of $\approx$0.27~\muHz{}. 

Frequency uncertainties were calculated using the method described by \citet{kb12}:
\begin{equation}
    \sigma(\nu) \approx 0.44 \sqrt{\pi/N}\sigma_{\rm noise} a^{-1} \sqrt{T_{\rm obs}^{-2} + \tau^{-2}},
\end{equation}
where $N$ is the number of points in the time series, $a$ is the mode amplitude, and $\tau$ is the mode lifetime. For the Sun, the mode lifetime is typically 1--3 days \citep{chaplin2000-lw,bedding2004-alp-cen}, and it increases for cooler stars. We assumed a conservative mode lifetime of 3 days for this star in the uncertainty calculation.

Table~\ref{table:freqs} lists the determined oscillation frequencies. Furthermore, we fitted Equation~\ref{eq:asymp} to these frequencies and obtained the following asymptotic parameters:
$\Dnu{} = 182.799 \pm 0.069$~\muHz{},
$\epsilon = 1.455 \pm 0.010$,
$\dnu{0,1} = 3.81 \pm 0.41$~\muHz{},
$\dnu{0,2} = 10.90 \pm 0.41$~\muHz{}, and 
$\dnu{0,3} = 22.27 \pm 0.59$~\muHz{}.

\section{Asteroseismic Modeling}\label{sec:modeling}

In this section, we derive the stellar parameters of HD~219134 using modern asteroseismic modeling techniques. 
We first describe the input observables used in the modeling process (\S\ref{subsec:obs}), followed by an overview of the modeling methodology and key assumptions (\S\ref{subsec:stp}). Finally, we provide brief comments on the derived stellar age and its implications (\S\ref{subsec:age}).

\subsection{Classical Constraints}\label{subsec:obs}

\begin{figure}
    \centering
    \includegraphics{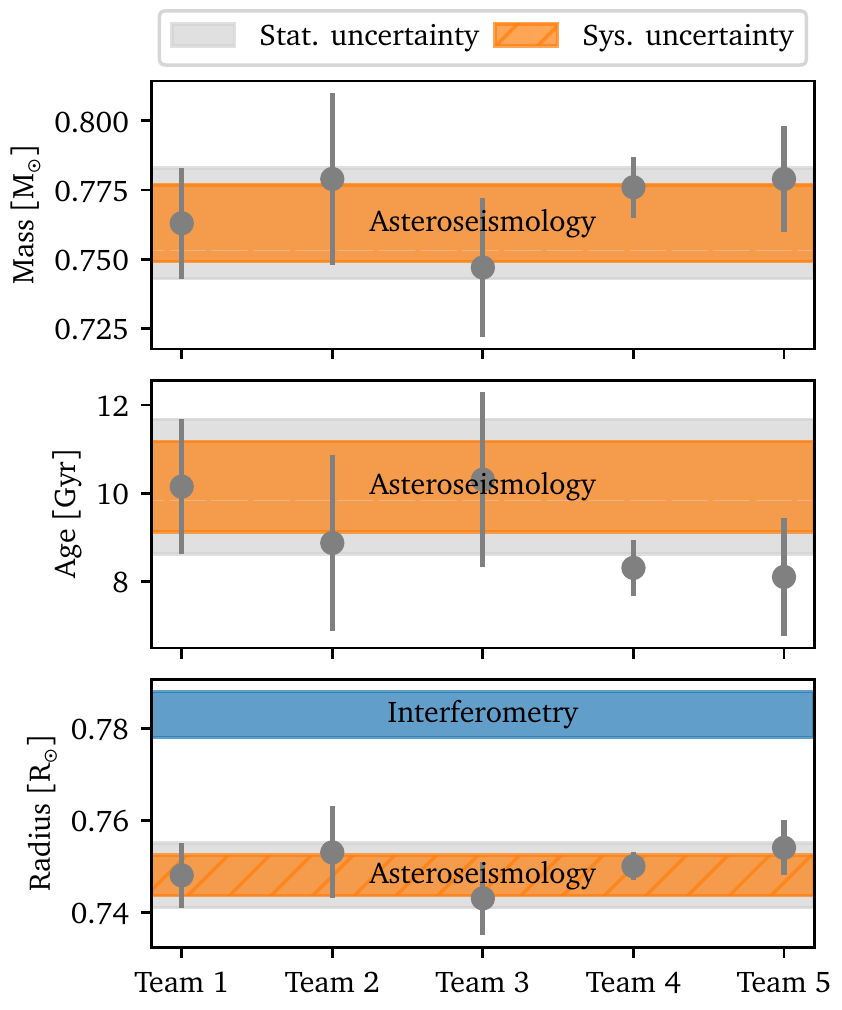}
    \caption{Stellar mass, radius and age derived from five independent asteroseismic modeling teams. The asteroseismic radius is compared against the interferometric radius based on an angular diameter measurement from \citet{elliott2025}. }
    \label{fig:summary}
\end{figure}

To derive a bolometric flux for HD~219134 we used optical broadband photometry from \textit{Tycho} \citep{hog2000-tyc}, since near-infrared 2MASS photometry is saturated and systematic errors for bright stars are not yet well-calibrated in Gaia \citep{riello21}.
The star has a $B_{\rm T}$ magnitude of $6.861\pm0.015$ and a $V_{\rm T}$ magnitude of $5.674\pm0.009$. Incorporating the Gaia DR3 distance and bolometric corrections, we used \texttt{ISOCLASSIFY} \citep{Huber2017,Berger2020} to calculate bolometric fluxes of $\fbol=(1.881 \pm 0.055)\times10^{-7}$~erg s$^{-1}$ cm$^{-2}$ and $\fbol=(1.981 \pm 0.054)\times10^{-7}$~erg s$^{-1}$ cm$^{-2}$, based on $B_{\rm T}$ and $V_{\rm T}$, respectively.
\citet{ligi2019-219134} determined the bolometric flux by fitting its spectral energy distribution, finding $\fbol=(1.986 \pm 0.024)\times10^{-7}$~erg s$^{-1}$ cm$^{-2}$. 
Due to the higher precision of the $V_{\rm T}$ magnitude, we adopted its corresponding bolometric flux value. To refine the uncertainty, we calculated the standard deviation of all three flux measurements to account for systematic uncertainties, and combined this in quadrature with the original uncertainty. 
This yielded a final value of $\fbol=(1.981\pm0.072)\times10^{-7}$~erg s$^{-1}$ cm$^{-2}$.

The star’s angular diameter was measured using CHARA/VEGA by \citet{ligi2019-219134} as $\theta=1.035\pm0.021$~mas, after correcting for limb darkening. Combined with the Gaia DR3 distance of $D=6.541\pm0.002$~pc \citep{bj2021-dist}, this results in a stellar radius of $R=0.728\pm0.015$~\Rsun{} via
\begin{equation}
    R=\frac{1}{2} \theta D.
\end{equation}

We calculated the star’s interferometric effective temperature based on the bolometric flux and angular diameter:
\begin{equation}
    \Teff=4^{1/4}f_{\rm bol}^{1/4}\sigma_{\rm SB}^{-1/4}\theta^{-1/2},
\end{equation}
where $\sigma_{\rm SB}$ is the Stefan-Boltzmann constant. 
The result is $\Teff{}=4854 \pm 66$~K, consistent with the spectroscopic temperature measured by \citet{rosenthal2021-cls}, who reported $\Teff = 4817 \pm 62$~K. The measured metallicity from \citet{rosenthal2021-cls} is $\feh{} = 0.083 \pm 0.058$~dex.

A recent measurement of the star’s angular diameter by \citet{elliott2025} is $\theta=1.114\pm0.007$~mas, which corresponds to a stellar radius of $R_\star = 0.783\pm0.015$~\Rsun{}, and an interferometric temperature of $\Teff{}=4678\pm45$~K. 
In principle, this angular diameter should be more accurate, given that the measurement spanned a wider range of spatial frequencies.
However, asteroseismic modeling using this interferometric \Teff{} predicts a radius that differs from the interferometric radius by $>3\sigma$.
This suggests the two methods may not be strictly compatible, resulting from either unaccounted interferometric systematics or inaccuracies in the stellar models (see \S\ref{sec:radius} for detail).

To maximise compatibility between inputs and stellar models, we adopted a conservative set of input observables for modeling: the interferometric \Teff{} based on \citet{ligi2019-219134}, spectroscopic [Fe/H] from \citet{rosenthal2021-cls}, and the oscillation frequencies presented in Table~\ref{table:freqs}. 
We deliberately excluded the bolometric flux and radius as direct constraints, despite their high precision, to avoid potential inconsistencies between these values and the asteroseismic constraints.
We summarize the stellar properties in Table~\ref{table:stp}.

\begin{deluxetable*}{cccccccc}
\tabletypesize{\scriptsize}
\tablecolumns{5}
\tablecaption{ Summary of the frequency modeling. \label{table:physics}}
\tablehead{
 & \colhead{Team 1} & \colhead{Team 2} & \colhead{Team 3} &  \colhead{Team 4} & \colhead{Team 5}\\
}
\startdata
\midrule
\multicolumn{6}{c}{\textbf{Model Configuration}}\\
\midrule
Stellar evolution code 
& MESA r240301 
& MESA r12778 
& GARSTEC v2015
& YREC 
& GARSTEC v2020\\
\midrule[0.1pt]
Pulsation code  
& GYRE v7.1 
& GYRE v5 
& ADIPLS v0.4 
& \citet{antia1994}
& GYRE v7.2\\
\midrule[0.1pt]
Opacities \& EOS 
& OPAL 
& OPAL 
& OPAL 
& OPAL
& OPAL\\
\midrule[0.1pt]
\makecell{Atmospheric\\boundary condition}
& \makecell{\citet{hauschildt1999a,hauschildt1999b} \\ \citet{castelli2003}}
& \citet{ks1966} 
& \citet{eddington1926} 
& \citet{eddington1926}
& \citet{trampedach2014} \\
\midrule[0.1pt]
Metal mixture 
& \citet{agss09} 
& \citet{agss09} 
& \citet{agss09} 
& \citet{gs98}
& \citet{gs98} \\
\midrule[0.1pt]
\makecell{Nuclear reaction rate}
& \makecell{\citet{jinareaclib2010} \\ \citet{nacre1999}} 
& \makecell{\citet{jinareaclib2010} \\ \citet{nacre1999}} 
& \makecell{\citet{nacre1999} \\ \citet{formicola2004} \\ \citet{hammer2005}} 
& \makecell{\citet{adelberger1998} \\ \citet{formicola2004}}
& \makecell{\citet{nacre1999} \\ \citet{formicola2004} \\ \citet{hammer2005}} \\
%
%
\midrule[0.1pt]
Mixing length formulation
& \citet{henyey1965} 
& \citet{cox1968} 
& \citet{bv1958} 
& \citet{bv1958}
& \citet{cox1968}  \\
\midrule[0.1pt]
\makecell{Mixing length parameter\\$\amlt$}
& Free 
& Solar-calibrated
& Free 
& Free 
& Varying vs. solar-calibrated \\
\midrule[0.1pt]
\makecell{$\alpha_{{\rm MLT},\odot}$}
& 1.96
& 1.71
& 1.79
& 1.84
& 1.83\\
\midrule[0.1pt]
Convective overshoot 
& \makecell{\citet{herwig2000}\\$f_{\rm ov, shell}=0.0174$} 
& None 
& None 
& None 
& None \\
\midrule[0.1pt]
Extra-mixing 
& None 
& \makecell{Gravitational settling\\Turbulent mixing}  
& \citet{thoul1994} 
& \citet{thoul1994}
& \makecell{\citet{thoul1994}\\Diffusive mixing}\\
\midrule[0.1pt]
Surface correction 
& \makecell{\citet{bg14} two-term\\+\citet{liyg2023} ensemble correction}  
& \citet{sonoi2015} 
& \makecell{\citet{bg14}\\cubic-term}
& \makecell{\citet{bg14}\\two-term}
& \makecell{\citet{bg14} two-term\\+ \citet{roxburgh_asteroseismic_2016} $\epsilon$-matching }  \\
\midrule[0.1pt]
Grid free parameters 
& $M$, [M/H], \amlt{}, \Yinit{}
& $M$, [M/H], \Yinit{} 
& $M$, [M/H], \amlt{}, \Yinit{}
& $M$, [M/H], \amlt{}, \Yinit{}
& $M$, [M/H], \Yinit{} \\
\midrule\midrule
\multicolumn{6}{c}{\textbf{Derived Stellar Parameters}}\\
\midrule
Mass $M_\star$ (\Msun{})
& $0.763\pm0.020$ 
& $0.779\pm0.031$ 
& $0.773\pm0.038$ 
& $0.776\pm0.011$
& $0.779\pm0.019$\\
Radius $R_\star$ (\Rsun{})
& $0.748\pm0.007$ 
& $0.753\pm0.010$
& $0.751\pm0.012$ 
& $0.750\pm0.003$
& $0.754\pm0.006$\\
Age $t_\star$ (Gyr)
& $10.2\pm1.5$  
& $8.9\pm2.0$ 
& $9.2\pm2.5$
& $8.3\pm0.6$
& $8.1\pm1.3$\\
Density $\rho_\star$ ($\rho_\odot$)
& $1.823\pm0.003$
& $1.825\pm0.001$
& $1.826\pm0.100$
& $1.824\pm0.001$
& $1.818\pm0.007$\\
Surface gravity $\logg_\star$ (dex)
& $4.577\pm0.001$
& $4.576\pm0.006$
& $4.570\pm0.005$
& $4.575\pm0.002$
& $4.575\pm0.004$\\
\Yinit{} 
& $0.270\pm0.023$  
& $0.278\pm0.022$
& $0.279\pm0.019$
& $0.280\pm0.005$
& $0.272\pm0.010$\\
$\amlt{}/\amlt_{,\odot}$ 
& $0.95\pm0.05$  
& ---
& $1.03\pm0.07$
& $0.94\pm0.04$
& ---\\
\enddata
\vspace{-0.8cm}
\end{deluxetable*}

\begin{deluxetable*}{lccccccc}
\tabletypesize{\footnotesize}
\tablecolumns{6}
\tablecaption{Star and planet properties.\label{table:stp}}
\tablehead{Property & Value & Reference }
\startdata
\multicolumn{3}{c}{\textbf{HD~219134}}\\
\midrule
\multicolumn{3}{c}{\textbf{Photometry}}  \\
\midrule
$B_T$-band magnitude & 6.861 $\pm$ 0.015 & \citet{hog2000-tyc} \\
$V_T$-band magnitude & 5.674 $\pm$ 0.009 & \citet{hog2000-tyc} \\
Bolometric flux \fbol{} (erg s$^{-1}$ cm$^{-2}$) & (1.981 $\pm$ 0.072) $\times10^{-7}$ & This work\\
\midrule
\multicolumn{3}{c}{\textbf{Astrometry}}  \\
\midrule
Distance (pc) & 6.5409 $\pm$ 0.0023 & \citet{bj2021} \\
Luminosity $L_\star$ (\Lsun{}) & 0.265 $\pm$ 0.011 & \S\ref{subsec:obs} \\ 
\midrule
\multicolumn{3}{c}{\textbf{Interferometry}}  \\
\midrule
Angular diameter $\theta$ (mas) & 1.035 $\pm$ 0.021 & \citet{ligi2019-219134} \\
Effective temperature \Teff{} (K) & 4854 $\pm$ 66 $^1$ & \S\ref{subsec:obs} \\
Radius $R_\star$ (\Rsun{}) & 0.728 $\pm$ 0.015 & \S\ref{subsec:obs} \\ 
\midrule
Angular diameter $\theta$ (mas) & 1.114 $\pm$ 0.007 & \citet{elliott2025} \\
Effective temperature \Teff{} (K) & 4678 $\pm$ 45 & \S\ref{subsec:obs} \\
Radius $R_\star$ (\Rsun{}) & 0.783 $\pm$ 0.005 & \S\ref{subsec:obs} \\ 
\midrule
\multicolumn{3}{c}{\textbf{Spectroscopy}} \\ 
\midrule
Effective temperature \Teff{} (K) & 4817 $\pm$ 62 & \citet{rosenthal2021-cls} \\
Metallicity \feh{} (dex) & 0.08 $\pm$ 0.06 $^1$ & \citet{rosenthal2021-cls} \\
\midrule
\multicolumn{3}{c}{\textbf{Asteroseismology}} \\ 
\midrule
\Dnu{} (\muHz{}) & $182.799 \pm 0.069$  & \S\ref{sec:obs} \\
$\epsilon$ & $1.455 \pm 0.010$ & \S\ref{sec:obs} \\
\dnu{0,1} (\muHz{}) & $3.81 \pm 0.41$ & \S\ref{sec:obs} \\
\dnu{0,2} (\muHz{}) & $10.90 \pm 0.41$ & \S\ref{sec:obs} \\
\dnu{0,3} (\muHz{}) & $22.27 \pm 0.59$ & \S\ref{sec:obs} \\
\numax{} (\muHz{}) & $4651 \pm 301$ & \S\ref{sec:modes} \\
\Aosc{} (cm~s$^{-1}$) & $4.23 \pm 0.41$ & \S\ref{sec:modes} \\
Mass $M_\star$ (\Msun{}) 
& 0.763 $\pm$ 0.020 (stat) $\pm$ 0.014 (sys) 
& \S\ref{subsec:stp} \\
Radius $R_\star$ (\Rsun{}) 
& 0.748 $\pm$ 0.007 (stat) $\pm$ 0.004 (sys) 
& \S\ref{subsec:stp} \\
Age $t_\star$ (Gyr) 
& 10.2 $\pm$ 1.5 (stat) $\pm$ 1.0 (sys) 
& \S\ref{subsec:stp} \\
Density $\rho_\star$ ($\rho_\odot$) 
& 1.823 $\pm$ 0.003 (stat) $\pm$ 0.003 (sys) 
& \S\ref{subsec:stp}\\
Surface gravity $\logg_\star$ (dex) 
& 4.577 $\pm$ 0.001 (stat) $\pm$ 0.003 (sys) 
& \S\ref{subsec:stp}\\
Initial helium abundance \Yinit{} 
& 0.27 $\pm$ 0.02 (stat) $\pm$ 0.01 (sys) 
& \S\ref{subsec:stp} \\
\midrule 
\multicolumn{3}{c}{\textbf{Rotation}} \\ 
\midrule
Rotation period \Prot{} (d) & 41.3 $\pm$ 2.8 & \S\ref{subsec:prot} \\
\midrule 
\multicolumn{3}{c}{\textbf{HD~219134b}} \\ 
\midrule
Mass (\Mearth{}) & 4.59 $\pm$ 0.16  & \S\ref{sec:planet} \\
Radius (\Rearth{}) &  1.542 $\pm$ 0.054 & \S\ref{sec:planet} \\
\midrule 
\multicolumn{3}{c}{\textbf{HD~219134c}} \\ 
\midrule
Mass (\Mearth{}) & 4.23 $\pm$ 0.20  & \S\ref{sec:planet} \\
Radius (\Rearth{}) & 1.455 $\pm$ 0.046  & \S\ref{sec:planet} \\
\enddata
\tablecomments{1. Used for asteroseismic modeling. }
\end{deluxetable*}

\subsection{Frequency modeling}\label{subsec:stp}

For frequency modeling, we employed five distinct modeling pipelines to assess the systematic uncertainties arising from different choices of input physics and the use of various codes.

Stellar evolution codes used by the five teams included MESA \citep{mesa2011,mesa2013,mesa2015,mesa2018,mesa2019,mesa2023,Moedas2024}, GARSTEC \citep{garstec2008}, and YREC \citep{yrec2008}. 
Pulsation codes used for calculating oscillation frequencies included GYRE \citep{gyre2013}, ADIPLS \citep{adipls2008}, and \citet{antia1994}. 

The input physics used by the five teams also differed.
Treatments of atmospheric boundary conditions included the Eddington \citep{eddington1926} and Krishna-Swamy \citep{ks1966} $T$-$\tau$ relations, and pre-computed photosphere tables \citep{hauschildt1999a,hauschildt1999b,castelli2003,trampedach2014}.
Choices of nuclear reaction rates varied from \citet{jinareaclib2010}, \citet{nacre1999}, \citet{formicola2004}, \citet{hammer2005}, and \citet{adelberger1998}.
The mixing length formulations included \citet{henyey1965}, \citet{cox1968} and \citet{bv1958}, with some teams using solar-calibrated mixing length parameters, others treating it as a free variable, and still others using calibrated prescriptions for varying it. 
Two main metal mixtures were used: those of \citet{agss09} and of \citet{gs98}.
Mass, metallicity, and initial helium abundance were all treated as free parameters by all teams.

Different teams employed different correction procedures for near-surface modeling errors. Several of the teams used empirical correction formulae \citep{bg14}, where an offset is applied to the mode frequencies to account for the effects of the surface term, before the corrected mode frequencies are used to construct a conventional likelihood function. These empirical formulae admit additional free parameters, which may either be fitted against the data, or potentially be externally calibrated against 3D models \citep{sonoi2015} or from an empirical ensemble \citep{liyg2023}. Other teams employed the nonparametric $\epsilon$-matching technique \citep{roxburgh_asteroseismic_2016}, where combinations of mode frequencies are computed from both the model and observed set, chosen so that the likelihood function evaluated from these combinations is insensitive to the near-surface structure of the star (see e.g. \citealt[][]{ong_surface_2021} for a review of surface-term correction techniques). We discuss in more depth in \S\ref{subsec:atm} their potential systematic effects on our stellar-modeling results.
Table~\ref{table:physics} lists the detailed configurations as well as the derived stellar parameters by each team.

Overall, the derived stellar properties for HD~219134 show strong consistency across modeling teams, despite intentional variations in input physics. The standard deviation of median values from each team is comparable to or smaller than the average uncertainties reported by each team, indicating that model uncertainties arising from differences in input physics and stellar evolution codes are generally smaller than the formal uncertainties reported.

We report our stellar properties of HD~219134 as follows: we adopted the median values and the statistical uncertainties from Team 1's median and formal uncertainty, and adopted the standard deviation of median values from each team as an estimate of the systematic uncertainties. 
The results are reported in Table~\ref{table:stp} and Figure~\ref{fig:summary}. 
We find a mass of 
$M_\star=\mass\pm\emass({\rm stat})\pm\smass({\rm sys})$~\Msun{}, a radius of 
$R_\star=\radius\pm\eradius({\rm stat})\pm\sradius({\rm sys})$~\Rsun{}, an age of 
$t_\star=\age\pm\eage({\rm stat})\pm\sage({\rm sys})$~Gyr, and an initial helium abundance of 
$\Yinit{}=0.27\pm0.02({\rm stat})\pm0.01({\rm sys})$.
The fractional uncertainties of mass, radius, and age are 3\%, 1\%, and 18\%, respectively. 

The derived \Yinit{} is close to the solar initial helium abundance ($\approx$0.27; \citealt{agss09}). Given that HD~219134’s metallicity is also close to solar, this \Yinit{} aligns well with expectations from Galactic enrichment laws \citep[e.g.,][]{lyttle2021}.
This constraint on \Yinit{} is derived largely indirectly from other constraints rather than directly from helium glitch signatures, as the latter typically produce signals below 1~\muHz{}, which is smaller than the statistical uncertainties reported in Table~\ref{table:freqs} \citep{verma2019-he, Mazumdar2014}.

Furthermore, we determine a stellar density of $\rho_\star=1.823\pm0.003 ({\rm stat})\pm0.003({\rm sys})$~$\rho_\odot$, and a surface gravity of $\logg_\star=4.577\pm0.001({\rm stat})\pm0.003 ({\rm sys})$~dex. The systematic uncertainties are comparable to or larger than the statistical uncertainties, highlighting the need to account for both sources of error for these properties \citep[e.g.][]{huber2022-20s}.

\begin{figure}
    \centering
    \includegraphics{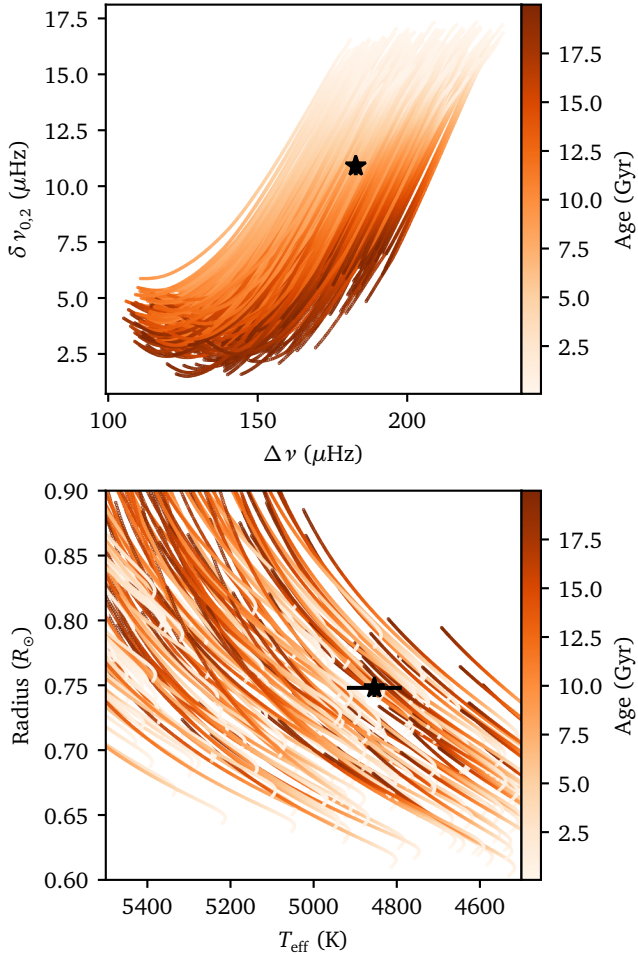}
    \caption{Stellar evolutionary models from zero-age to terminal-age main sequence, calculated by Team 1, presented on the C--D diagram (top; \dnu{0,2} vs. \Dnu{}) and the H--R diagram (bottom). The models are color-coded by stellar age. The position of HD~219134 is marked by a star symbol, with 1$\sigma$ uncertainties, some of which are smaller than the symbol size.}
    \label{fig:age}
\end{figure}

\subsection{Asteroseismic Age is Less Affected by Surface Physics}\label{subsec:age}

Our estimate of the asteroseismic age ($t_\star=\age\pm\eage({\rm stat})\pm\sage({\rm sys})$~Gyr) is approximately 1-$\sigma$ lower than the age derived by \citet{gillon2017-219134}, who reported $t_\star = 11.0 \pm 2.2$~Gyr based on modeling that incorporated a precise interferometric radius.

Unlike constraints based on radius, asteroseismic ages primarily depend on constraints on the star’s internal structure, particularly the hydrogen-burning processes near the core. 
Figure~\ref{fig:age} presents stellar evolutionary models computed by Team 1 alongside the star’s position. On the H–R diagram, stellar tracks with different input parameters overlap at the star’s position, making it difficult to determine the stellar age based solely on surface properties without implementing any priors on the input parameters. However, the C--D diagram provides a clearer age diagnostic, as \dnu{0,2} traces the fraction of hydrogen burned in the core \citep{jcd1984-02}. 

Stellar properties derived using asteroseismic constraints is less influenced by assumptions in surface modeling or the accuracy of surface constraints such as radius and \Teff{}.
For example, Teams 1 and 3 examined the effect of different choices of classical observables, either using \Teff{} alone or in combination with the interferometric radius from \citet{ligi2019-219134}, along with asteroseismic constraints. They found that the resulting ages remained consistent within the quoted uncertainties.

\section{Radius Inflation}\label{sec:radius}

In Table~\ref{table:stp} and Figure~\ref{fig:summary}, we compare the asteroseismic radius derived in this study with the interferometric radius determined by \citet{elliott2025}. 
The asteroseismic radius derived for HD~219134, $R_\star=\radius\pm\eradius({\rm stat})\pm\sradius({\rm sys})$~\Rsun{}, is significantly lower than the new interferometric radius, $R_\star = 0.783\pm0.005$~\Rsun{} \citep{elliott2025}.
Both radii are highly precise, with uncertainties below 1\%, yet they differ by approximately 4\% (a 4$\sigma$ discrepancy).

While the exact cause of this discrepancy is still unclear, in this section, we explore several explanations. 
Future observations of more targets common to both asteroseismology and interferometry could help identify the source of this discrepancy.

\subsection{Systematic Uncertainties in Interferometry}

Interferometric visibilities can be affected by systematic errors due to assumed calibrator diameters, wavelength calibration, or limb-darkening corrections \citep[e.g.][]{vb05}. Systematic differences in interferometric angular diameters measured using different methods and instruments can reach up to 10\% \citep{white18,tayar2022}. These systematic errors are most significant if the angular diameter of a target star is not well resolved compared to the calibrator diameters, which is supported by systematic trends between photometric and interferometric temperature scales \citep{casagrande14}. 

\citet{ligi2019-219134} used the optical VEGA beam combiner \citep{m11} on the CHARA Array \citep{mcalister05} to measure the angular diameter of for HD~219134 using baselines up to $\approx$\,100\,m, yielding a minimum visibility of $V^2 \approx 0.2-0.3$. The latest interferometric data obtained using the optical PAVO beam combiner on the CHARA Array \citep{ireland08} used a longer baseline up to $\approx$\,160\,m, thus better resolving the star down to the first visibility null \citep{huber2016-itf, elliott2025}. This makes the derived angular diameter essentially independent of systematics due to assumed calibrator sizes. Systematic effects due to limb darkening corrections and wavelength calibration are expected to be well below the 1\% level \citep{huber-phd,huber2012}. Given previous good agreement between PAVO and VEGA for stars that are similarly well resolved \citep{ligi12, white13}, we conclude that it is unlikely that the $\approx$\,4\% radius tension between interferometry and asteroseismology can be attributed to systematic errors in the interferometric measurements. Systematic errors in parallax scales by Gaia at this level can also be ruled out \citep{Groenewegen2021}.

\subsection{Atmospheric Boundary Conditions}
\label{subsec:atm}

We note that the majority of the asteroseismic constraints presented in this work are explicitly constructed to be insensitive to the surface (and therefore radius) of the star. This is because modeling errors --- arising from the neglect of more realistic equilibrium structures in the stellar atmosphere, and the interaction between convection and oscillations --- are known to produce a systematic offset between frequencies of actual stars, and those of notional 1D stellar models with identical internal structures. Corrections for this so-called ``surface effect" must therefore be applied to model frequencies, in order to mitigate this discrepancy, but doing so often renders these asteroseismic constraints insensitive to the structure of the near-surface layers through the introduction of additional free parameters. Under these circumstances, the reported radius does not depend significantly on the precise form chosen to describe this surface effect using free parameters \citep{nsamba_surface_2018,ong_surface_2021}. 

Conversely, this means that the radius estimates returned from these asteroseismic modeling efforts are primarily model-dependent, in that they are the radii of 1D models possessing interior structures consistent with the asteroseismic constraints, rather than radii directly constrained using seismology. All else being equal, the radii of 1D models in turn are determined by the atmospheric boundary conditions imposed on them computationally.

We confirmed that the determined radius is not highly sensitive to some specific modeling choices. This was demonstrated by modifying only one aspect of the models while keeping all other physical assumptions fixed.
To assess the influence of the atmospheric boundary condition, Team 1 replaced the photosphere tables \citep{hauschildt1999a,hauschildt1999b,castelli2003} with the \citet{eddington1926} model, resulting in a slight radius decrease of 0.006~\Rsun{}. Team 4 examined the effect of replacing \citet{eddington1926}’s $T$-$\tau$ relation with that of \citet{ks1966} and found a small radius decrease of 0.009~\Rsun{}. Additionally, in implementing the surface correction procedure, Team 2 replaced their use of the calibrated \citet{sonoi2015} formulation (which has no free parameters) with that of \citet{bg14} (which includes free parameters), leading to a minor radius reduction of 0.001~\Rsun{}.

Moreover, the modeling teams employed a diverse set of atmospheric models, including $T$-$\tau$ relations such as those of Eddington \citep{eddington1926} and Krishna-Swamy \citep{ks1966}, and precomputed atmosphere tables based on more realistic 1D or 3D models \citep{hauschildt1999a,hauschildt1999b,castelli2003,trampedach2014}. Despite this broad range of approaches, the differences in the predicted radius across the modeling teams remained small, which suggests that the common choice of atmospheric boundary conditions and surface correction procedure fails to explain the radius discrepancy.

\subsection{Mixing Lengths}\label{subsec:amlt}

\begin{figure}
    \centering
    \includegraphics{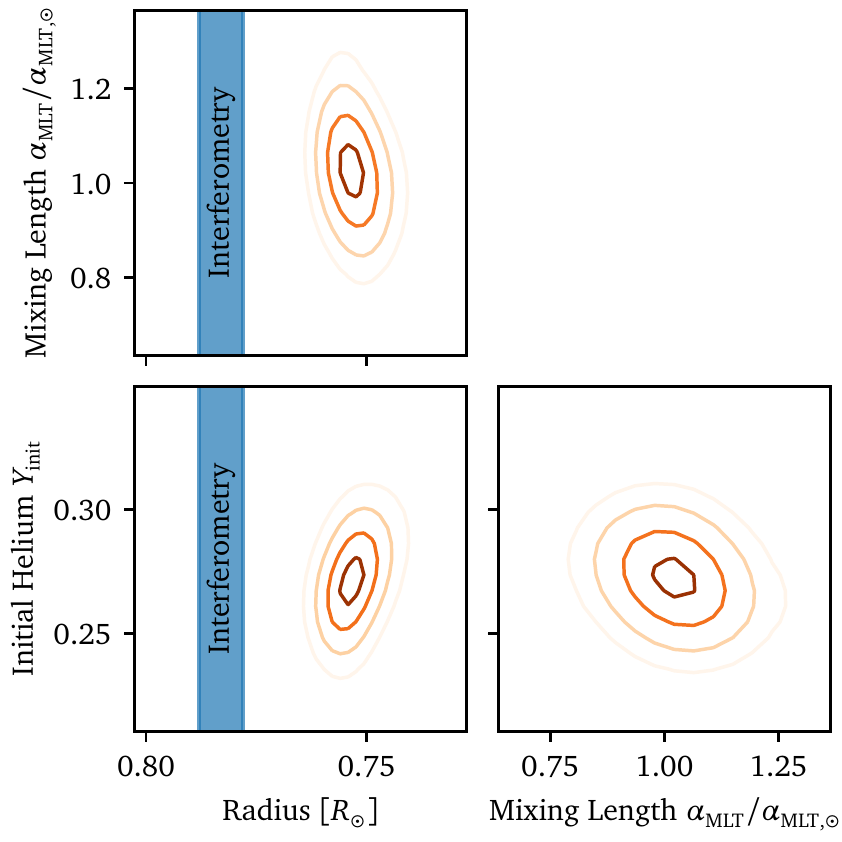}
    \caption{Probability contours of stellar properties presented in the space of radius, \amlt{} and \Yinit{}, based on models calculated by Team 1. These contours are derived from constraints provided by asteroseismic observations. }
    \label{fig:corner}
\end{figure}

The radii of 1D stellar models with convective envelopes are determined by convective mixing-length theory, which specifies the temperature stratification of the superadiabatic layers close to the stellar surface. Since most of our modeling teams treated the mixing length parameter \amlt{} either as a global free parameter or a fixed one according to calibrations against the Sun, incorrect values of \amlt{} might have systematically biased the resulting asteroseismic model radius constraints.

3D hydrodynamical simulations produce entropy profiles associated with this temperature stratification that cannot be reproduced in 1D stellar evolution when \amlt{} is treated as a globally constant quantity. Calibrated prescriptions for \amlt{} also exist \citep[e.g.][]{spada_stellar_2021,trampedach2014}, anchored to relations between the adiabatic entropy, surface gravity, and effective temperature that are seen to emerge in these simulations \citep{Magic2015,tanner_entropy_2016}. One modeling team (Team 5) performed their modeling using the variable-\amlt{} prescription of \cite{trampedach2014}, but the radius and age scale of this modeling was still ultimately determined by a solar-calibrated value of $\amlt = 1.83$, against which the local $\amlt$ was obtained by scaling under this prescription.

There have also been suggestions that \amlt{} should change as a function of stellar mass and/or metallicity in main-sequence dwarfs. 3D simulations indicate that the entropy jump between the bottom and top of the super-adiabatic region is smaller for cooler dwarfs \citep{Trampedach2013}. This trend corresponds to higher convective efficiency and, consequently, a larger \amlt{} \citep{Magic2015}.
These suggest that $\amlt{}/\alpha_{\text{MLT},\odot}$ at 4800~K is between 0.95 and 1.10 \citep{trampedach2014,Magic2015}, consistent with 2D simulations using gray radiative transfer \citep{Ludwig1999}.

1D stellar models, when constrained by oscillation frequencies and atmospheric parameters, can also provide estimates of \amlt{} \citep[e.g.,][]{Viani2018}. Observations of the well-characterized visual binary system $\alpha$ Cen A and B further support the trend of increasing \amlt{} with decreasing stellar mass \citep[][]{joyce2018-acen}, even though the fit of \amlt{} in 1D models allows the compensation for other deficiencies in the underlying model physics. 

These studies all support the notion that \amlt{} at 4800~K should be near or slightly above solar. 
Accordingly, we investigated the shift of radius as a response to changing \amlt{}.
Figure~\ref{fig:corner} presents the posterior distributions of radius, \amlt{}, and initial helium abundance (\Yinit{}) using Team 1’s stellar models. 
We observe a correlation between \amlt{} and radius from the distributions. However, allowing a 20\% change in \amlt{} from the best-fitting value still fails to reconcile the predicted radius with that derived from interferometry. Furthermore, larger values of \amlt{} reduce \Yinit{} to values near or below the primordial helium abundance of 0.249 \citep{Planck2016}, which is physically implausible.

Furthermore, the use of different input physics (investigated by this paper) is unlikely to shift \amlt{} significantly. We see near-solar \amlt{} across modeling teams where \amlt{} was treated as a free parameter: averaging around $\amlt/\amlt_{\odot} = 0.980\pm0.053$. 
We also found that the \amlt{} values are not significantly affected by changes in their boundary conditions. 
This is supported by comparisons between models that differ only in their boundary conditions: Team 1 repeated their analysis, replacing their photosphere tables with the \citet{eddington1926} boundary condition, and Team 4 replaced the \citet{eddington1926} boundary condition with that of \citet{ks1966}. 
Both teams found the changes in $\amlt{}/\amlt{}_\odot$ remained within their quoted 1$\sigma$ confidence intervals.

\subsection{Magnetic Fields}
K and M dwarfs are often observed to have larger radii than those predicted by stellar models, a phenomenon referred to as radius inflation \citep{torres2013}. 
This discrepancy can be resolved in models by incorporating magnetic fields directly in stellar structure equations \citep{feiden2012,macdonald2012,feiden2013}, or by including starspots that inhibit energy transport on the stellar surface \citep{Somers2015}.
Observational evidence suggests that the degree of radius inflation correlates with surface magnetic field strength \citep{kiman2024} and rotation period \citep{lanzafame2017-rad}, supporting the role of magnetic activity as a key factor.

Modeling studies found that strong magnetic fields are necessary to account for radius inflation.
\citet{Somers2015} found that a 50\% spot coverage is required to explain a 4\% radius difference (see their Figure 2). \citet{feiden2013} studied UV Psc B, which shows a 10\% radius inflation compared to non-magnetic models, requiring surface magnetic fields of approximately 4~kG to match the observations. 

However, HD~219134 appears not to possess a magnetic field strong enough to explain the observed discrepancy. 
The surface magnetic field strength measured by \citet{folsom2018-zdi} is on the order of 2.5~G, comparable to the Sun and the $\alpha$ Cen A and B systems \citep{feiden2012}. 
Furthermore, observations of the star’s chromospheric emission using Ca II H and K lines spanning over a decade (see \S\ref{subsec:activity}) places HD~219134 firmly in the inactive regime of stellar activity distribution. Thus, magnetic activity is very unlikely to account for the 4\% difference between the asteroseismic and interferometric radii.

\subsection{Tidal Heating}

The orbital angular momentum of a planet can drive tidal interactions within the convective envelope of its host star, leading to energy dissipation in the form of heat. This dissipation can act as an additional energy source, potentially contributing to the inflation of the host star. However, in the case of HD~219134, which hosts two ultra-short-period super-Earths, HD~219134b and HD~219134c, the expected tidal effect is insufficient to account for a 4\% increase in the stellar radius.

To examine this, we calculated the tidal decay timescale, $\tau_d$, for a slowly rotating host star using the following expression  \citep{Winn2018,Goldreich1966,Dai2024}:
\begin{equation}
    \tau_d \approx 30 ~\text{Gyr} 
    \left(\frac{Q'_\star}{10^6}\right) \left(\frac{M_\star/\Msun}{M_p/\Mearth} \right)
    \left(\frac{\rho_\star}{\rho_\odot}\right)^{5/3}
    \left(\frac{P_{\rm orb}}{1~\text{day}}\right)^{13/3},
\end{equation}
where $Q'_\star$ is the tidal quality factor, which we assume to be $10^7$ \citep{Penev2018}, and $P_{\rm orb}$ is the orbital period.
Using this timescale, we derived the corresponding energy dissipation rate:
\begin{equation}
    \dot{E} = \frac{3}{2} \frac{G M_\star M_p}{P_{\rm orb}} \frac{1}{\tau_d}.
\end{equation}
With the stellar and planetary parameters from Table~\ref{table:stp} and \citet{gillon2017-219134}, we determined that the energy dissipation induced by the close-in planet HD~219134b is $1.17 \times 10^{-7}~\Lsun{}$ . 
Even if all this dissipated energy were converted into stellar luminosity, it would still fall several orders of magnitude short of explaining the required 8\% luminosity discrepancy (corresponding to the 4\% difference in radius).

\section{Evolution of Rotation and Activity}\label{sec:rotation}

Using the estimated asteroseismic age, $t_\star=10.2\pm1.5({\rm stat})\pm1.0({\rm sys})$~Gyr, alongside an independently measured rotation period, we can test the theory of angular momentum loss.

\begin{figure}
    \centering
    \includegraphics{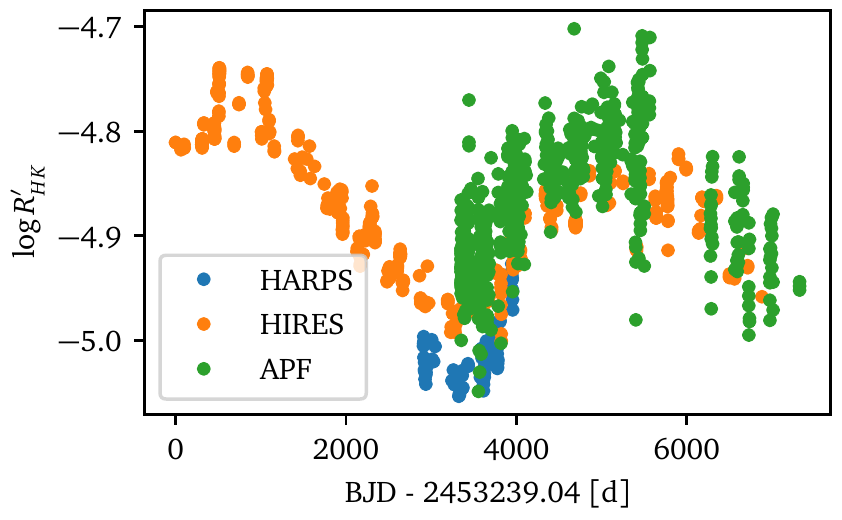}
    \caption{Time series of \logrphk{} measured by three different instruments spanning 20 years (HARPS, \citealt{motalebi2015-219134}; HIRES and APF, \citealt{rosenthal2021-cls,isaacson2024-cls-cycle}).}
    \label{fig:activity-ts}
\end{figure}

\begin{figure*}
    \centering
    \includegraphics{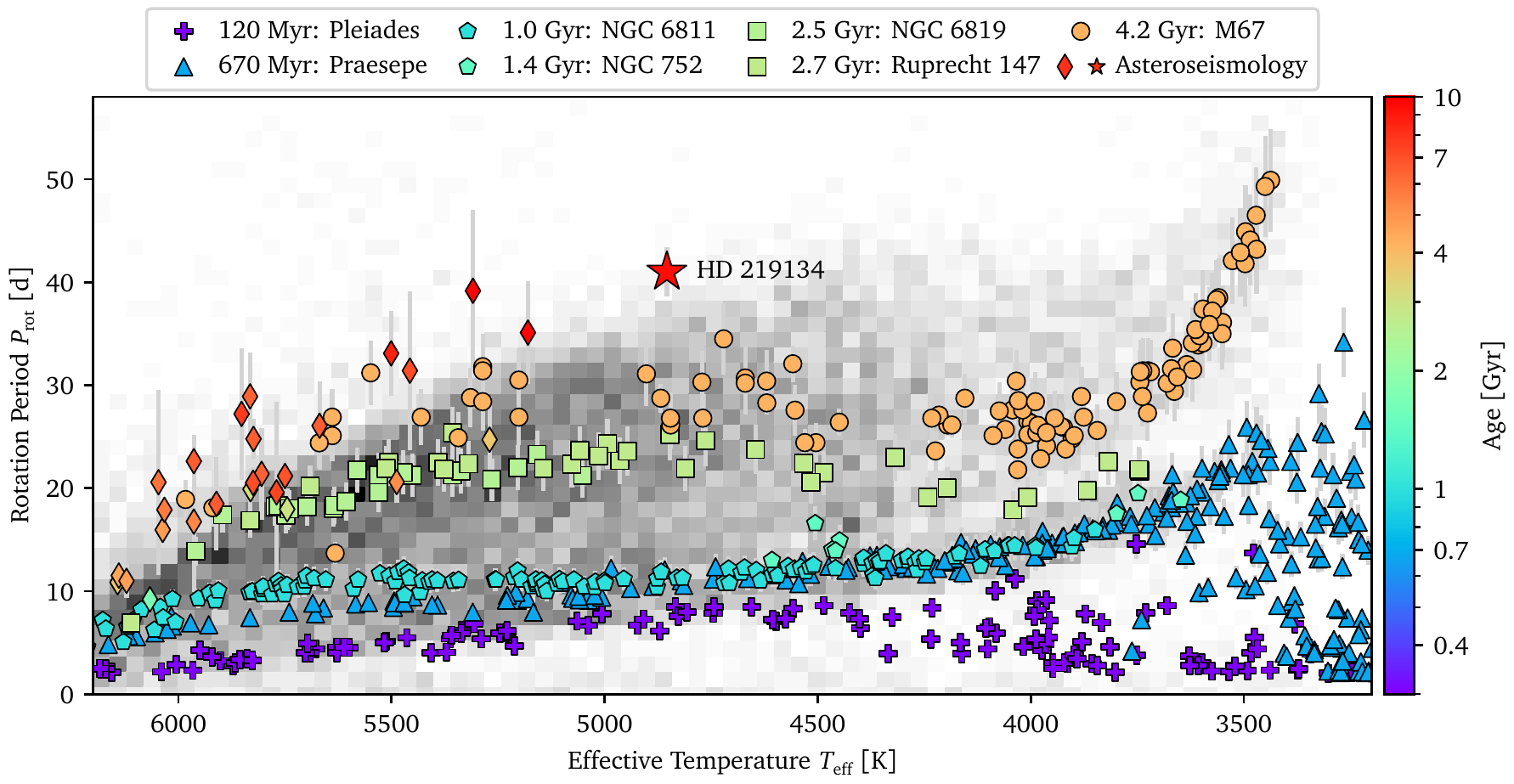}
    \caption{Calibrators of gyrochrones shown on the \Prot{}--\Teff{} diagram. The open cluster data include Pleiades \citep{rebull2016-pleiades}, Praesepe \citep{douglas2017-praesepe,douglas2019-praesepe}, NGC 6811 \citep{curtis2019-6811}, NGC 752 \citep{agueros2018-752}, NGC 6819 \citep{meibom2015-6819}, Ruprecht 147 \citep{curtis2020-147}, and M67 \citep{barnes2016-m67,dungee2022-m67,long2023-cl}. The asteroseismic field stars shown include stars with both rotation periods and ages measured via asteroseismology \citep{hall2021,silvaaguirre2015-kages,silvaaguirre2017-legacy}. The background displays a binned distribution of the \Kepler{} field star sample \citep{mcquillan2014}. HD~219134 is the coolest age calibrator for stars older than 4.2~Gyr.}
    \label{fig:gyro_teff}
\end{figure*}

\begin{figure}
    \centering
    \includegraphics{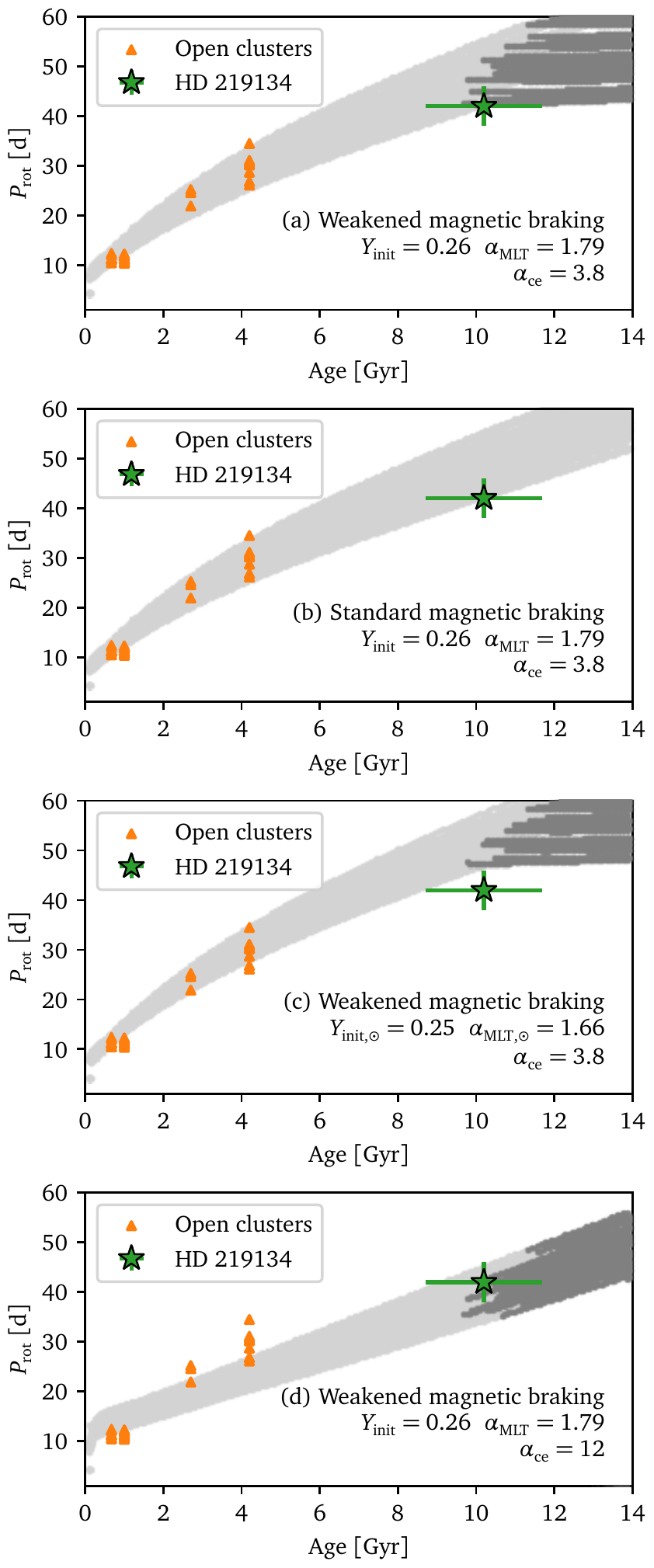}
    \caption{Rotation evolutionary models projected to older ages. These are calibrated based on asteroseismic field stars and open clusters hotter than 4500 K. The calibrators within the $\Teff\in[4650,4900]$~K range are shown in triangles. The evolutionary tracks include models within 3-$\sigma$ that match HD~219134’s mass and metallicity. Additional model configurations are specified in each panel. Models highlighted in dark gray represent those that have entered the weakened magnetic braking phase ($\dd J_{\rm tot} / \dd t=0$).
    }
    \label{fig:gyro_age}
\end{figure}

\subsection{Rotation Period}\label{subsec:prot}

Based on Calcium H and K line emission measured with HARPS-N over three years, \citet{motalebi2015-219134} reported a rotation period for HD~219134 of $\Prot=43.2$~d. Similarly, \citet{folsom2018-zdi}, using Zeeman–Doppler imaging from Narval spectropolarimeter, found a rotation period of $\Prot=42$~d. 
The spread in rotation period estimates from different methods is consistent with expected effects of latitudinal differential rotation.

We used HIRES and APF data collected via California Planet Search \citep{rosenthal2021-cls,isaacson2024-cls-cycle}, as well as HARPS data \citep{motalebi2015-219134} to determine the star's rotation period. 
These data are shown in Figure~\ref{fig:activity-ts}. 
We modeled the time series using a hierarchical model and Gaussian processes \citep{celerite2}. Each time series was assigned a single rotation period, with all periods assumed to follow a normal distribution with a common mean.
As a result, we determined a rotation period of $\Prot=41.3\pm2.8$~d.

Figure~\ref{fig:gyro_teff} shows the position of this star on the \Prot{}--\Teff{} plane, compared with other stars of known rotation period and benchmark ages derived from open clusters or asteroseismology. 
Stellar angular momentum loss over time is driven primarily by magnetized stellar winds. Establishing a well-calibrated rotation-age relationship is essential for age-dating cool dwarf stars, especially for exoplanet hosts. 
HD~219134 is noteworthy for occupying a unique parameter space as the coolest benchmark star with an age exceeding 4.2 Gyr.

\subsection{Rotation Modeling}\label{subsec:rotevol}

To check whether our current knowledge of magnetic braking and angular momentum transport is sufficient to explain the rotation period of HD~219134 at $\age$ Gyr, we used the stellar evolution models calculated from Team 1, and calculated stellar rotation using \texttt{ROTEVOL} \citep{vanSaders2013-sg,somers2017}.

Models were initialized with a rotation period of $P_{\rm disk}=4$~d and disk-locking timescale of $\tau_{\rm disk}=10$~Myr, following the choices of \citet{chiti2024-wd}. We employed a magnetic braking model based on \citet{vanSaders2016-wmb}, which quantifies the total angular momentum loss as:
\begin{equation}\label{eq:amloss}
    \dv{J_{\rm tot}}{t} = \left\{ 
    \begin{aligned}
        & f_K K_M \omega_e \left( \frac{\omega_{\rm sat}}{\omega_\odot} \right)^2, \ 
        \omega_{\rm sat} \leq \omega_e \frac{\tau_{\rm cz}}{\tau_{\rm cz,\odot}}, 
        \Ro{} \leq \Ro_{\rm crit}  \\
        & f_K K_M \omega_e \left( \frac{\omega_e\tau_{\rm cz}}{\omega_\odot \tau_{\rm cz,\odot}} \right)^2, \ 
        \omega_{\rm sat} > \omega_e \frac{\tau_{\rm cz}}{\tau_{\rm cz,\odot}},
        \Ro{} \leq \Ro_{\rm crit}  \\
        & 0, \ \Ro > \Ro_{\rm crit} 
    \end{aligned}
    \right.
\end{equation}
Here, $\omega_e$ is the envelope rotation rate, $\tau_{\rm cz}$ is the convective overturn timescale, calculated locally at one pressure scale height above the bottom of the outer convection zone, $\omega_{\rm sat}=3.863\times10^{-5}$~rad~s$^{-1}$ is the saturation threshold of angular momentum loss, \Ro{} is the Rossby number defined as $\Prot/\tau_{\rm cz}$ as a means to estimate stellar magnetism, and $\Ro_{\rm crit}$ is a Rossby threshold, beyond which no angular momentum loss occurs \citep{vanSaders2016-wmb,saunders2024-ro}. 
The factor $K_M$ scales with stellar properties as: 

\begin{equation}
    \frac{K_M}{K_{M,\odot}} = c
    \left(\frac{R}{R_\odot}\right)^{3.1} \\ \left(\frac{M}{M_\odot}\right)^{-0.22}
    \left(\frac{L}{L_\odot}\right)^{0.56} \left(\frac{P_{\rm phot}}{P_{\rm phot,\odot}}\right)^{0.44},
\end{equation}
where $c$ corrects for centrifugal force (assumed to be 1), and $P_{\rm phot}$ denotes the photospheric pressure.

We modeled the stellar rotation using a two-zone model, with the core and envelope rotating independently as solid bodies, and exchanging angular momentum over a coupling timescale $\tau_{\rm ce}$. This timescale scales with stellar mass as $\tau_{\rm ce}/\tau_{\rm ce, \odot} = (M/M_{\odot})^{-\alpha_{\rm ce}}$, with $\tau_{\rm ce, \odot}\approx22$~Myr \citep{sl20}. The angular momentum loss rates for the core and envelope are expressed as \citep{denissenkov2010-ce}
\begin{equation}
    \dv{J_c}{t} = -\frac{(\Delta J)_{\rm max}}{\tau_c} - \omega_s \dv{I_e}{t},
\end{equation}
\begin{equation}
    \dv{J_e}{t} = \frac{(\Delta J)_{\rm max}}{\tau_c} + \omega_s \dv{I_e}{t} + \dv{J_{\rm tot}}{t},
\end{equation}
where
$I_e$ and $I_c$ represent envelope and core inertia, respectively.
The term $\omega_s$ is defined as the envelope rotation rate $\omega_e$ if the envelope is contracting, or the core rotation rate $\omega_c$ if the envelope is expanding.
The maximum angular momentum exchange between the core and envelope is given by:
\begin{equation}
    (\Delta J)_{\rm max} = \frac{I_c I_e}{I_c + I_e}(\omega_c-\omega_e).
\end{equation}

The free parameters in this model are $f_K$, $\Ro_{\rm crit}/\Ro_\odot$ and $\alpha_{\rm ce}$. To calibrate these parameters, we constructed a sample of open clusters, field stars, and the Sun with their rotation periods and ages. The open clusters were the Pleiades \citep[120 Myr;][]{rebull2016-pleiades}, Praesepe \citep[670 Myr;][]{douglas2017-praesepe,douglas2019-praesepe}, NGC 6811 \citep[1 Gyr;][]{curtis2019-6811}, NGC 6819 \citep[2.5 Gyr;][]{meibom2015-6819}, Ruprecht 147 \citep[2.7 Gyr;][]{curtis2020-147}, and M67 \citep[4.2 Gyr;][]{barnes2016-m67,dungee2022-m67,long2023-cl}. The field stars were \Kepler{} stars with rotation periods \citep{hall2021} and asteroseismic ages \citep{silvaaguirre2015-kages,silvaaguirre2017-legacy}. 
Finally, the Sun was included as a data point with $\Prot{}=25.4$~d and $t_\odot=4.67$~Gyr \citep{Bahcall1995}. 
We restricted the sample to stars with effective temperatures above 4500~K, to avoid difficulties in modeling stars around 4200~K, which have enhanced activity that is thought to be due to shear-enhanced magnetism \citep[e.g.][]{cao2023}.
The best-fitting model parameters obtained are $f_K=5.655$, $\Ro_{\rm crit}/\Ro_\odot=0.93$, and $\alpha_{\rm ce}=3.8$.

\subsection{Testing Rotation Models Beyond 4.2~Gyr}\label{subsec:gyro}

We searched these rotational evolutionary tracks for models with the same mass and metallicity as HD~219134 by predicting rotation periods to the age of HD~219134.

Figure~\ref{fig:gyro_age}(a) presents these evolutionary tracks, calculated using the default weakened magnetic braking configuration. 
These models assume initial helium abundance (\Yinit{}) and mixing length parameter (\amlt{}) constrained by asteroseismology, along with the best-fitting core-envelope coupling coefficient ($\alpha_{\rm ce}$).
The models agree very well with HD~219134’s observed rotation period and age. The estimated $\Ro/\Ro_\odot$ is 0.83$\pm$0.07, only slightly below the critical value $\Ro_{\rm crit}/\Ro_\odot = 0.93$, indicating the star is around the transition point into the weakened magnetic braking regime. 

\citet{isaacson2024-cls-cycle} reported an activity cycle of $\Pcyc = 13.27$~yr for HD~219134, placing it on the lower sequence in the \Pcyc{}–\Prot{} diagram. Following \citet{metcalfe2022-wmb}, this positioning also implies HD~219134 is close to the transition into weakened magnetic braking \citep{vanSaders2016-wmb}, consistent with our findings.

In fact, \citet{Metcalfe2025} found evidence of significantly weakened instantaneous torque in HD~219134, based on constraints on magnetic field strength and morphology from ZDI, suggesting the star has already entered the weakened magnetic braking phase. Given that our estimate of $\Ro/\Ro_\odot$ is $1\sigma$ below $\Ro_{\rm crit}/\Ro_\odot$, this provides tentative (though not statistically significant) evidence that $\Ro_{\rm crit}$ for this star may be different from most G-dwarf asteroseismic calibrators.

We also modeled rotational evolution assuming standard magnetic braking, which prescribes continued spin-down even when $\Ro > \Ro_{\rm crit}$ (modifying the third line in Equation~\ref{eq:amloss} to equal the second line). The results are shown in Figure~\ref{fig:gyro_age}(b). As expected, this model predicts further spin-down beyond 10~Gyr. 
If the observation of instantaneous torque is not considered, both the standard and weakened braking models adequately explain the current observations. This is because the divergence in rotation periods between the two regimes requires a longer timescale to become clearly distinguishable.

Next, we tested whether models with solar-calibrated \Yinit{} and \amlt{} could also explain the observations. These results are shown in Figure~\ref{fig:gyro_age}(c). Compared with Figure~\ref{fig:gyro_age}(a), models with \Yinit{} and \amlt{} constrained by asteroseismology provide a better overall fit. The initial helium abundance (\Yinit{}) affects the stellar structure by altering the size of the convection zone, and hence $\tau_{\rm cz}$ \citep{saunders2024-ro}, similar to the effects of [Fe/H] or [$\alpha$/M] \citep{Claytor2020}. This highlights the importance of incorporating detailed chemical abundances for accurately predicting ages using gyrochronology.

Finally, we explored models with a large coupling timescale motivated by studies such as \citet{cao2023}, who found $\alpha_{\rm ce}\approx12$ for late K dwarfs with enhanced magnetism in the Praesepe open cluster. These models are presented in Figure~\ref{fig:gyro_age}(d). While the rotation period and age of HD~219134 agree with the model, the rotation period for stars around 3 Gyr is underestimated. Thus, our data do not support a larger coupling timescale for stars at this \Teff{}. This suggests that the scaling relation for coupling timescale versus mass may be oversimplified for modeling the stalled spin-down phase.

In summary, weakened magnetic braking models with constrained abundances provide a good match to HD~219134’s observed rotation period and asteroseismic age. Our findings support the reliability of such models for predicting rotational evolution in early K-type dwarfs beyond 4.2~Gyr. 
The effects of prolonged core-envelope coupling earlier in stellar evolution can leave observable imprints in older stars. Combining asteroseismic ages and torque measurements for a statistically significant sample will help constrain the coupling timescale as a function of \Teff{}.

\subsection{Activity-Age Relations}\label{subsec:activity}

\begin{figure}
    \centering
    \includegraphics{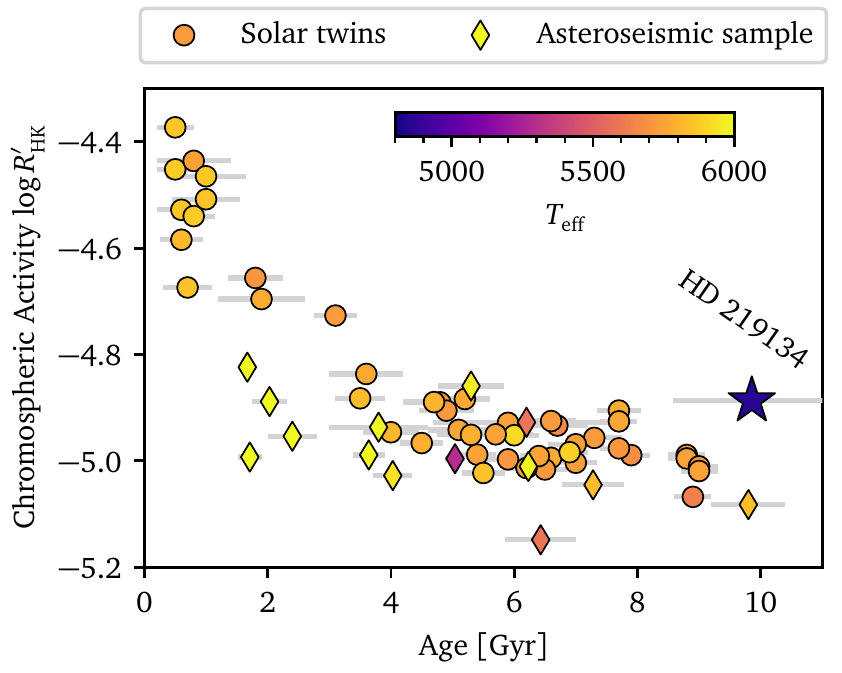}
    \caption{Chromospheric activity-age relation. The sample includes solar twins \citep{lo2018}, the Kepler asteroseismic sample \citep{karoff2013,metcalfe2016,creevey2017}, and the TESS asteroseismic sample \citep{huber2022-20s,chontos2021,metcalfe2020-94aqr,metcalfe2021-rcrb}.}
    \label{fig:activity}
\end{figure}

A common method for age-dating field stars involves using activity-age relations. These activity indices include metrics such as spot modulation amplitudes, emissions in spectral lines (e.g., Calcium H\&K and H$\alpha$), and emissions in the X-ray or radio spectrum. In this study, we examined whether the chromospheric activity of HD~219134 aligns with established activity-age relationships.

Wherever possible, we used the color-temperature relation from \citet{Sekiguchi2000} and converted the S-indices into the color-independent activity index \logrphk{}, on the scale defined by \citet{Middelkoop1982} and \citet{Noyes1984}.
Figure~\ref{fig:activity} presents \logrphk{} for a sample of solar twins \citep{lo2018} and asteroseismic samples from Kepler and TESS \citep{karoff2013,metcalfe2016,creevey2017,huber2022-20s,chontos2021,metcalfe2020-94aqr,metcalfe2021-rcrb}. 
We found that at an age of 10~Gyr, HD~219134 exhibits a higher activity level compared to solar twins of the same age but is comparable to the activity levels observed in solar twins at 5~Gyr. 
This is consistent with expectations that activity levels scale with \Ro{}, as \Ro{} of HD~219134 is close to $\Ro_{\rm crit}$, a value also observed for solar twins at 5~Gyr, beyond which they enter into the phase of weakened magnetic braking.
This result supports the inclusion of a \Teff{} dependence in the \logrphk{}–age relation, particularly for cooler stars, as differences in \Ro{} can affect the activity-age trend by as much as 5~Gyr.

\section{Revising Planet Properties}\label{sec:planet}

\begin{figure}
    \centering
    \includegraphics[width=\columnwidth]{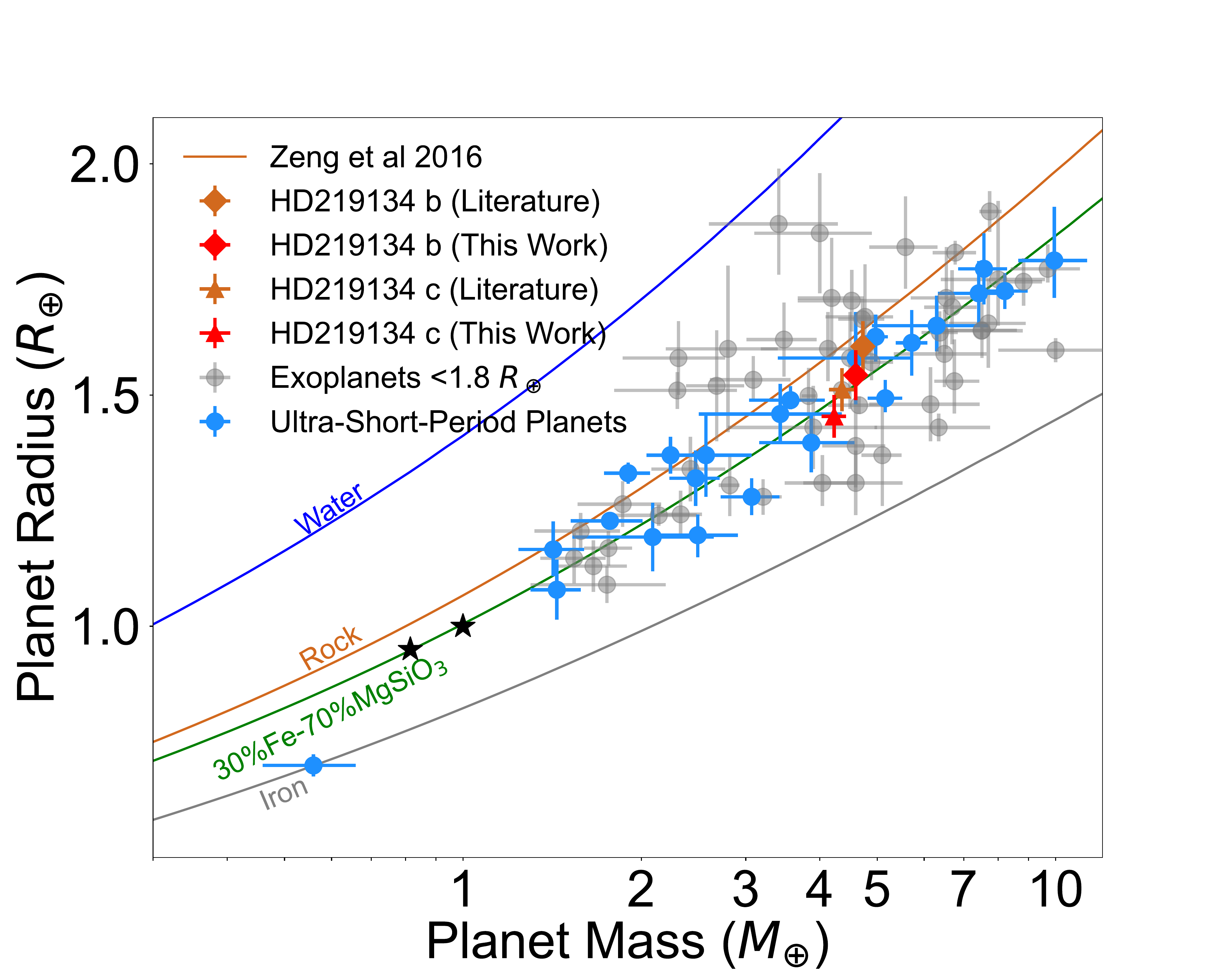}
    \caption{The mass--radius diagram of exoplanets with $R_p<1.8~\Rearth$, using planet properties from the NASA Exoplanet Archive \citep{NASAExoplanetScienceInstitute2020}. Ultra-short-period planets are defined following \citet{dai2019}, and are expected to be bare rocky cores whose composition can be probed directly with mass and radius measurements. The solid lines indicate different theoretical compositional models calculated by \citet{Zeng2016}. }
    \label{fig:planet}
\end{figure}

HD~219134 hosts five confirmed exoplanets, including two transiting super-Earths (HD~219134b and HD~219134c) first discovered by \citet{gillon2017-219134} from photometric analysis using the Spitzer Space Telescope. As one of brightest star known to host transiting planets, the system has been extensively studied, including the possibility of the formation and detection of planetary exospheres \citep{Vidotto18}.

We produce revised properties of the two transiting planets based on our new asteroseismic stellar mass and radius. The mass of transiting planets can be calculated using
\begin{equation}
    M_p = \frac{K_1^2 + \sqrt{K_1^4 + 4 G(1-e^2)^{-1} \sin^2 i \ a^{-1} K_1^2 M_\star}}{2 G(1-e^2)^{-1} \sin^2 i \ a^{-1}},
\end{equation}
where $K_1$, $e$, $i$, $a$, $G$ denote the RV semi-amplitude, orbital eccentricity, orbital inclination, semi-major axis, and the gravitational constant, respectively.
The radius of a transiting planet is given by
\begin{equation}
    R_p = R_\star \cdot R_p' / R_\star',
\end{equation}
where $ R_p' / R_\star'$ is the old planet and star radius ratio determined by \citet{gillon2017-219134}.

Using the orbital parameters provided in \citet[][see their Table 1]{gillon2017-219134} along with the asteroseismic mass and radius derived in this work, we revise the HD~219134b's mass and radius from 
$M_p = 4.73\pm0.17$~\Mearth{} and $R_p = 1.605\pm0.055$~\Rearth{} 
to $M_p = 4.59\pm0.16$~\Mearth{} and $R_p = 1.542\pm0.054$~\Rearth{}. Similarly, we revise the mass and radius of HD~219134c from
$M_p = 4.36\pm0.21$~\Mearth{} and $R_p = 1.512\pm0.046$~\Rearth{} 
to $M_p = 4.23\pm0.20$~\Mearth{} and $R_p = 1.455\pm0.046$~\Rearth{}. 

We plot both the previous and updated planet properties on the mass-radius diagram in Figure~\ref{fig:planet}. The revised mass and radius shift the original data points slightly downward, supporting the hypothesis that ultra-short-period super-Earths typically exhibit an Earth-like composition of 30\% Fe and 70\% MgSiO$_3$ composition \citep{dressing2015,dai2019,brinkman2024}.

\section{Oscillation Amplitudes}\label{sec:modes}

\begin{figure}
    \centering
    \includegraphics{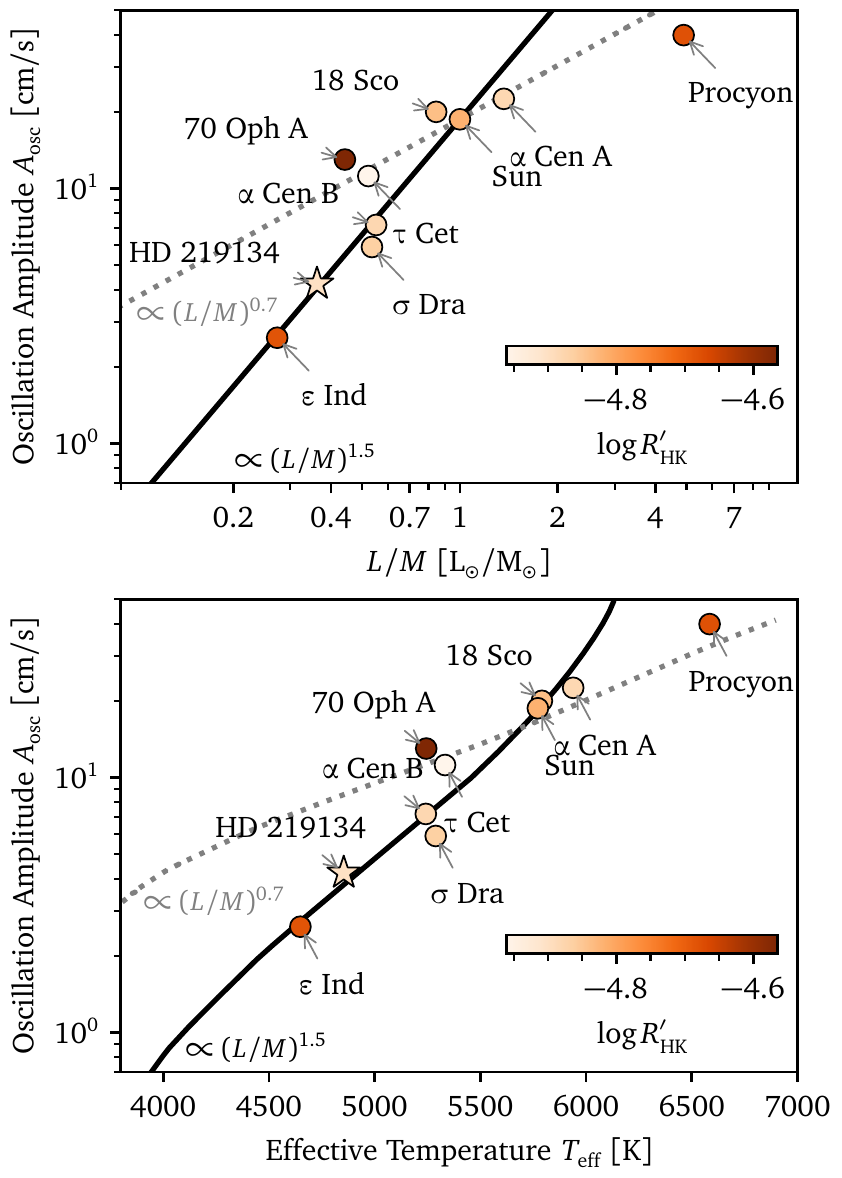}
    \caption{Oscillation amplitudes in radial velocity as a function of $L/M$ (top) and \Teff{} (bottom) for main-sequence dwarfs, color-coded by chromospheric activity.}
    \label{fig:amplitude}
\end{figure}

Using \texttt{pySYD}, we measured the mode amplitude \Aosc{} and the frequency of maximum power \numax{} from a heavily smoothed power spectrum. We obtained $\Aosc=4.23\pm0.41$~cm/s, and $\numax=4651\pm301$~\muHz{}.

The power spectrum of HD~219134 appears to show a dip in the center of the oscillation region that is remarkably similar to that seen in $\alpha$~Cen~B, which the only other K dwarf with a high-SNR power spectrum (see Figure 4 of \citealt{kjeldsen2005-acenb}). 
Similar features are also seen in $\epsilon$~Ind \citep[Figure 3 of][]{campante2024-eps-indi}, and $\sigma$~Dra \citep[Figure 6 of][]{hon2024-sig-dra}, though based on shorter time series. 
It remains to be seen whether this is a real feature of K dwarfs, which would be important for our understanding of the excitation and damping mechanisms of solar-like oscillations \citep[see review by][]{Houdek+Dupret2015}.

Figure~\ref{fig:amplitude} places HD~219134 among a sample of cool dwarf stars with oscillation amplitudes measured from radial velocity.
The pulsation amplitude in radial velocity is theoretically expected to scale as $\Aosc \propto (L/M)^s$, with $s$ ranging from 0.7 to 1.5 \citep{jcd1983,kb95,houdek1999,samadi2007}. 
Figure~\ref{fig:amplitude} also overlays MIST isochrones \citep{Choi2016} to illustrate the relationship between \Aosc{} and \Teff{}, assuming \Aosc{} scales with $(L/M)$ at power indices of 0.7 and 1.5, evaluated at ages of 1 and 4.6~Gyr, respectively.
While an $s$ index of 0.7 is consistent with observations of G- and F-type stars, \citet{campante2024-eps-indi} first suggested that cooler stars, such as $\epsilon$ Indi, align more closely with $s\approx$1.5. Our analysis of HD~219134 supports this result.

It has also been proposed that higher stellar activity can reduce oscillation amplitudes. However, despite HD~219134 and $\tau$ Ceti being relatively inactive, they follow the $s\approx$1.5 relation, indicating that activity alone is unlikely to account for the steeper amplitude scaling observed in cooler stars.

\section{Discussion and Conclusion}\label{sec:conc}

In this work, we present an asteroseismic analysis of HD~219134 using 4 consecutive nights of radial velocity data obtained with Keck Planet Finder on Keck-I. 
We used the Gold deconvolution algorithm to disentangle the power spectrum from the spectral window, in order to permit identification of modes separated by close to $1\ \mathrm{c/d}$, coinciding with the characteristic spacing of sidelobes. A total of 25 oscillation modes with spherical degrees $0\le\ell\le3$ were identified and extracted, as summarized in Table~\ref{table:freqs}. We then estimated HD~219134's properties using five independent asteroseismic modeling pipelines. The resulting parameters are: mass \mass{} $\pm$ \emass{} (stat) $\pm$ \smass{} (sys) \Msun{}, radius \radius{} $\pm$ \eradius{} (stat) $\pm$ \sradius{} (sys) \Rsun{}, and age \age{} $\pm$ \eage{} (stat) $\pm$ \sage{} (sys) Gyr. These in turn led to revised masses and radii of the known transiting planets orbiting HD~219134, as detailed in Table~\ref{table:stp}.

Asteroseismology with EPRV instruments is a promising avenue for calibrating gyrochronology and studying angular momentum loss in old K and M dwarfs. 
HD~219134 in particular is the first dwarf cooler than 5000~K with an asteroseismic age estimate available. We used this to test models of rotational evolution used for this purpose by confronting them, at this age, with an independent measurement of its present rotational period. 
We demonstrated that existing models of angular momentum loss, incorporating weakened magnetic braking and asteroseismically constrained values of \Yinit{} and \amlt{}, accurately reproduce HD~219134’s rotation period at its asteroseismic age. 
We further found that the spin-down relation for stars at this temperature requires a shorter spin-down coupling timescale than those suggested for late K dwarfs.

Nonetheless, we note that this agreement, and ultimately the apparent robustness of our asteroseismic age estimate, requires that our evolutionary stellar modeling be accurate. Such an assumption of accuracy in modeling also fundamentally underpins our asteroseismic radius constraints, since these are (as we discuss in \S\ref{subsec:atm}) the radii of stellar models consistent with asteroseismic constraints on stellar interiors, rather than stellar radii directly constrained using asteroseismology. Importantly, we have also found our asteroseismic radius to be 4\% smaller than the interferometric radius at 4-$\sigma$ level, possibly indicating some deficiency in the accuracy of this modeling. We were unable to easily attribute this discrepancy to any systematic uncertainties related to interferometry, nor to variations in canonical choices of atmospheric boundary conditions or mixing length theory used in stellar modeling, nor to magnetic fields, nor tidal heating. Without any insight into the cause of this discrepancy, our subsequently derived quantities, and treatment of rotational evolution --- all of which are contingent on these model ages and radii --- must necessarily be regarded as being only conditional, pending a better understanding of the physical origin for this discrepancy.
Future direct constraints on stellar radii from asteroseismology (e.g. through potential breakthroughs in understanding and mitigating the surface term) may alleviate this dependence on evolutionary modeling.

Finally, we have confirmed that HD~219134's oscillation mode amplitudes scale with those of other cool dwarfs as $(L/M)^{1.5}$, deviating from the $(L/M)^{0.7}$ relation observed in G-type stars. Should this difference in scaling behaviour be genuine and general, this may have important implications for the interpretation of other asteroseismic observables, namely $\numax$, in similar stars. In particular, both the oscillation amplitudes and $\numax$ are thought to be fully determined by the properties of turbulent stresses in the near-surface layers --- both ultimately reflect the physical processes that excite normal modes to visible amplitudes in the first place. If the convective turbulence that operates G-type stars should indeed be fundamentally different in some fashion from in cooler dwarfs --- yielding oscillation amplitudes that scale differently in these two classes of stars --- then we might also expect a different $\numax$ scaling relation to be required for these cooler dwarfs. Suggestively, HD~219134 joins $\alpha$~Cen~B \citep{kjeldsen2005-acenb}, $\epsilon$~Ind \citep{campante2024-eps-indi}, and $\sigma$~Dra \citep{hon2024-sig-dra}, in appearing to exhibit a broad excitation envelope that has a local minimum in mode amplitudes close to $\numax$, differing qualitatively from the solar phenomenology of a single, largely symmetric, envelope. Further, more detailed analyses of non-adiabatic pulsations and convective excitation, and a larger observational sample of likewise cool dwarfs, will be needed both to determine the exact cause of this difference, as well as to determine whether the solar-calibrated $\numax$ scaling relation indeed also holds for stars with modified amplitude scaling.

\section*{Acknowledgements}
The authors wish to recognize and acknowledge the very significant cultural role and reverence that the summit of Maunakea has always had within the Native Hawaiian community. We are most fortunate to have the opportunity to conduct observations from this mountain.
Y.L. acknowledges support from Beatrice Watson Parrent Fellowship and a NASA Keck PI Data Award, which is administered by the NASA Exoplanet Science Institute. Data presented herein were obtained at the W. M. Keck.
Observatory from telescope time allocated to the National Aeronautics and Space Administration through the agency's scientific partnership with the California Institute of Technology and the University of California. The Observatory was made possible by the generous financial support of the W. M. Keck Foundation.
D.H. acknowledges support from the Alfred P. Sloan Foundation, the National Aeronautics and Space Administration (80NSSC22K0781), and the Australian Research Council (FT200100871).
J.M.J.O. acknowledges support from NASA through the NASA Hubble Fellowship grant HST-HF2-51517.001, awarded by STScI. STScI is operated by the Association of Universities for Research in Astronomy, Incorporated, under NASA contract NAS5-26555.
T.R.B. acknowledges support from the Australian Research Council through Laureate Fellowship FL220100117.
T.L.C. is supported by Funda\c c\~ao para a Ci\^encia e a Tecnologia (FCT) in the form of a work contract (CEECIND/00476/2018).
M.S.L. is supported by a research grant (42101) from VILLUM FONDEN.
N.S. acknowledges support by the National Science Foundation Graduate Research Fellowship Program under Grant Numbers 1842402 \& 2236415 and NASA’s Interdisciplinary Consortia for Astrobiology Research (NNH19ZDA001N-ICAR) under award number 19-ICAR19 2-0041.

\appendix
\section{Radial Velocity Data}
In Table~\ref{table:rvs}, we list the full radial velocity data measured from the KPF spectra obtained in this work. The raw KPF spectra can be accessed from Keck Observatory Archive.

\begin{deluxetable}{cccccccc}
\tabletypesize{\footnotesize}
\tablecolumns{3}
\tablewidth{\textwidth} 
\tablecaption{ Radial velocity data of HD~219134 measured from KPF spectra. \label{table:rvs}}
\tablehead{
\colhead{\hspace{0.8cm}BJD$-2460579.72$ (d)}\hspace{0.8cm} & \colhead{\hspace{0.8cm}RV (m$\cdot$s$^{-1}$)}\hspace{0.8cm} & \colhead{\hspace{0.8cm}RV error (m$\cdot$s$^{-1}$)}\hspace{0.8cm}  \\
}
\startdata
0.00946579 & -18372.4687 & 0.4382 \\
0.01016011 & -18371.5709 & 0.4500 \\
0.01085437 & -18371.2015 & 0.4398 \\
0.01154667 & -18372.0286 & 0.4438 \\
0.01222188 & -18371.9946 & 0.4491 \\
0.01292862 & -18371.1237 & 0.4518 \\
0.01360698 & -18371.4315 & 0.4257 \\
0.01429799 & -18371.9583 & 0.4265 \\
0.01500247 & -18371.9011 & 0.4265 \\
0.01567058 & -18371.1193 & 0.4272 \\
... & ... & ... \\
\enddata
\tablecomments{Only the first 10 lines are shown. The full table can be accessed online.}
\end{deluxetable}

\bibliography{yaguang,bedding,others}{}
\bibliographystyle{aasjournal-compact}




\end{CJK}
\end{document}